\documentclass[twocolumn]{autart}    

\usepackage{amsmath}
\usepackage{algorithm}
\usepackage{algpseudocode}
\usepackage{amssymb}
\usepackage{color}
\usepackage{enumerate}
\usepackage{wrapfig}
\usepackage{graphicx}          
\usepackage[dvips]{epsfig}    
\usepackage[round, sort]{natbib}
\usepackage{hyperref}
\hypersetup{
	colorlinks = true,
	citecolor = cyan,
}
\newtheorem{lemma}{Lemma}[section]
\newtheorem{theorem}{Theorem}[section]

\newtheorem{remark}{Remark}[section]
\newtheorem{definition}{Definition}[section]

\newtheorem{assumption}{Assumption}[section]

\allowdisplaybreaks
\begin{document}
\begin{frontmatter}
	
	\title{Self-triggered Resilient Stabilization of Linear Systems	with Quantized Output} 
	
	\thanks[footnoteinfo]{The work was supported in part by the National Key R\&D Program of China under Grant 2021YFB1714800, the National Natural Science Foundation of China under Grants 62088101, 62173034, 61925303, 
		and the Chongqing Natural Science Foundation under Grant 2021ZX4100027. }
	\thanks[footnoteinfo]{This paper was not presented at any IFAC 
		meeting. }
	
	\author[Bit,Tcic]{Wenjie Liu}\ead{liuwenjie@bit.edu.cn}, 
	\author[Kobe]{Masashi Wakaiki}\ead{wakaiki@ruby.kobe-u.ac.jp},   
	\author[Bit,Tcic]{Jian Sun}\ead{sunjian@bit.edu.cn},            
	\author[Bit,Tcic]{Gang Wang}\ead{gangwang@bit.edu.cn}, 
	\author[Bit,Tongji]{Jie Chen}\ead{chenjie@bit.edu.cn} 
	
	\address[Bit]{Key Lab of Intelligent Control and Decision of Complex Systems, Beijing Institute of Technology, Beijing 100081, China}  
	\address[Tcic]{Beijing Institute of Technology Chongqing Innovation Center, Chonqing 401120, China}     
	\address[Kobe]{Graduate School of System Informatics, Kobe University, Hyogo 657-8501, Japan}             
	\address[Tongji]{Department of Control Science and Engineering, Tongji University, Shanghai 201804, China}        


	\maketitle

	\begin{abstract}
		This paper studies the problem of stabilizing a self-triggered control system with quantized output. Employing a standard observer-based state feedback control law, a self-triggering mechanism that dictates the next sampling time based on quantized output is co-developed with an output encoding scheme. If, in addition, the transmission protocols at the controller-to-actuator (C-A) and sensor-to-controller (S-C) channels can be adapted, the self-triggered control architecture can be considerably simplified, leveraging a delicate observer-based deadbeat controller to eliminate the need for running the controller in parallel at the encoder side.
		To account for denial-of-service (DoS) in the S-C channel, the proposed output encoding and self-triggered control schemes are further made resilient. It is shown that a linear time-invariant system can be exponentially stabilized if some conditions on the average DoS duration time are met.
	There is a trade-off between the maximum inter-sampling time and the resilience against DoS attacks.
		Finally, a numerical example is presented to demonstrate the practical merits of the proposed self-triggered control schemes and associated theory.
	\end{abstract}
\end{frontmatter}	
	\section{Introduction}~\label{sec:intro}
	With the development of communication and networking technologies, networked control systems (NCSs), in which data are transmitted over wired or wireless networks, have been widely integrated in modern engineering systems 
	(e.g., \cite{FP-RC-FB:10p,chen2021From}).
	In the study of NCSs, a 
	basic problem relates to determine how frequently different devices (e.g., sensors, controllers, and actuators) of a control system shall be executed so as to balance between the communication cost and the overall system performance. Transmission frequency and limited bandwidth are two major factors affecting the communication cost, both of which have been considerably studied in past decades.
	
	To reduce transmission frequency, research on aperiodic sampling techniques has lately aroused great interest.
		Among many,	event-triggered control and self-triggered control schemes are two practically appealing solutions
		\cite{heemels2012introduction}.
		In the former, the state or output of a plant is periodically or
		continuously sampled to examine the event-triggering condition, but the sampled value is transmitted only when the condition is met, e.g.,  \cite{aaarzen1999simple,heemels2013model,
				Wakaiki2020event}.
		In self-triggered control systems however, measurements are taken and transmitted only at sampling times which are determined based on previous sampled measurements, e.g., \cite{anta2010to,gommans2015resource,matsume2020resilient,Wakaiki2021selftrigger}.
		Hence, both sensors and communication channels are activated only at sampling times.	
		In general, self-triggered control outperforms event-triggered control
		in terms of prolonging the sensor lifetime, and reducing the communication load at the sensor-to-controller (S-C) channel.
	
	On the other hand, data transmitting through a channel with limited bandwidth are quantized before sent out.
	It has been shown that coarse quantization can deteriorate the system performance, and may even lead to instability \cite{bullo1576851}.
To secure system stability, the quantizer should be carefully designed.
Proposed in \cite{Liberzon2000Quantized}, the so-called ``zooming-in'' and ``zooming-out'' method offers an elegant way to design dynamic quantizers for linear time-invariant (LTI) systems.
Its generalizations can be found to deal with, e.g., nonlinear systems in  \cite{Liberzon2005nonlinear}, switched systems in \cite{Wakaiki2017Stabilization}, 
and systems under DoS attacks in \cite{8880482,Liu2021resilient}.

	Although it is natural to consider self-triggering mechanism and constrained bandwidth simultaneously in the real scenario,
	design a self-triggering mechanism using only coarse measurements at sampling times 
	is challenging.
	For this reason, it has not been fully addressed in the literature, and only few works have jointly studied these two factors; see e.g., 
	\cite{persis2013self,zhou2018self,Wakaiki2021selftrigger,ikeda2016quantized,liu2023data}.
	In \cite{persis2013self}, consensus protocols considering both self-triggered and quantized communication were proposed for linear multi-agent systems.
	Although limited bandwidth was considered in \cite{ikeda2016quantized,zhou2018self}, both works designed the self-triggering mechanisms using non-quantized states, which has addressed most of the difficulties encountered in self-triggered control.
	Most recently, the work \cite{Wakaiki2021selftrigger} proposed a self-triggering mechanism based on quantized states.
	
	The goal of this paper is to generalize the results of \cite{Wakaiki2021selftrigger} to systems where only quantized output rather than quantized state are available. Specifically, we consider that the controller-to-actuator (C-A) channel is ideal with adequate bandwidth, whereas the S-C channel has limited communication resources.
	The S-C channel hosts an encoder, a decoder, and a self-triggering module that are responsible for when/how the output shall be sampled, quantized, and transmitted.	
	In this context, a self-triggering mechanism and an output encoding scheme are co-designed for a linear system with a standard observer-based controller.
	In addition, if a higher transmission rate can be afforded at the C-A channel, 
	an observer-based deadbeat control law is designed such that the self-triggering mechanism can be considerably simplified and the computational overhead at the encoder can be reduced. 
	The key lies in synchronizing the design of the deadbeat controller gain and the transmission period of the C-A channel, in the use of the system controllability index.
		It is worth remarking that, under the proposed deadbeat controller, the self-triggering mechanism for linear systems with quantized outputs is as simple as that with quantized states in \cite{Wakaiki2021selftrigger}.
	To account for DoS attacks, a resilient self-triggered control scheme with quantized output is put forth, for which exponential system stability is established under certain conditions on the DoS duration time.
	We also reveal an intriguing trade-off between the maximum inter-sampling time and the system resilience in the presence of DoS attacks.
	As the self-triggering parameter decreases to yield the minimum inter-sampling time, the DoS condition in this paper coincides with that for the time-triggered sampling in \cite{8880482}.
	
In succinct form, the main contributions of this work are summarized as follows. 
		\begin{itemize}
			\item[\textbf{c1)}]
			To reduce the communication cost, an output encoding scheme and a self-triggering mechanism are co-developed for LTI systems by employing a standard observer-based control law;
			\item[\textbf{c2)}]
			When a higher communication rate can be afforded at the C-A channel, a delicate observer-based deadbeat control scheme significantly simplifies the self-triggering design, which is as simple as that for systems with quantized state feedback control in the literature;
			and,
			\item[\textbf{c3)}]
			Finally, the proposed self-triggered control and the encoding scheme are further made resilient against DoS attacks, and a trade-off between the maximum inter-sampling time and the system resilience is unraveled.
		\end{itemize}
	
Denote the set of real numbers by $\mathbb{R}$.
Given $\alpha \in \mathbb{R}$, let $\mathbb{R}_{>\alpha}$ 
($\mathbb{R}_{\ge \alpha}$) 
denote the set of real numbers greater than 
(greater than or equal to) $\alpha$.
Let $\mathbb{N}$ denote the set of natural numbers and define $\mathbb{N}_0 := \mathbb{N} \cup \{0\}$.
	For a vector $v = [v_1, v_2, \cdots\!, v_n]^T \in \mathbb{R}^n$, denote its maximum norm by $\Vert v\Vert_{\infty} := \max\{|v_1|, \cdots\!, |v_n|\}$ and the corresponding induced norm of a matrix $M \in \mathbb{R}^{m \times n}$ by $\Vert M\Vert_{\infty} := \sup\{\Vert Mv\Vert_{\infty}:v \in \mathbb{R}^n, \Vert v\Vert_{\infty} = 1\}$.
	
	\section{Preliminaries and Problem Formulation} \label{problem_Formulation}
\begin{figure}[t]
		\centering
		\includegraphics[width=7cm]{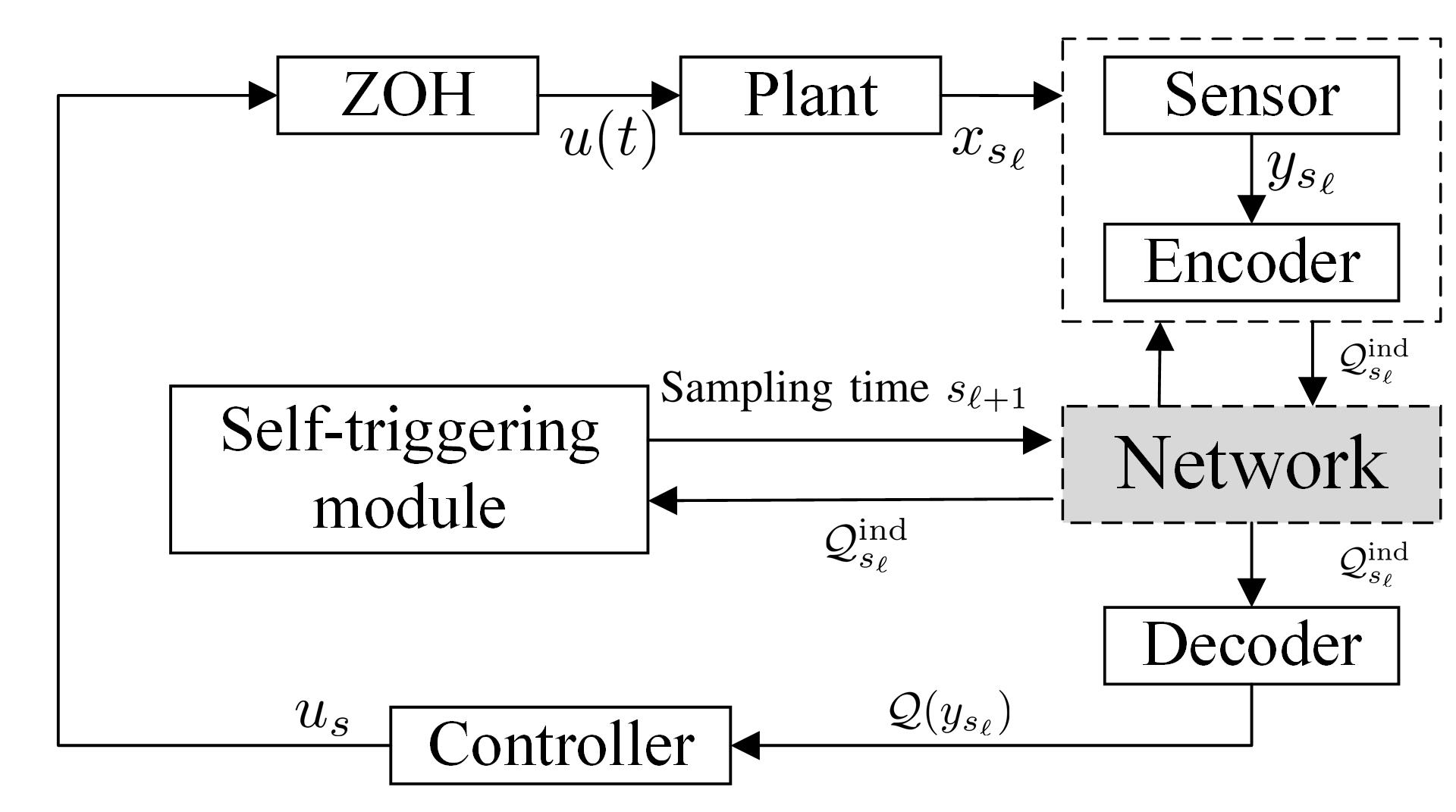}\\
		\caption{Closed-loop system with self-triggered control.}\label{fig:sys}
		\centering
	\end{figure}

	In this paper, we study the networked control architecture in Fig. \ref{fig:sys}, where a plant is to be stabilized 
	based on quantized output transmitted over a bandwidth-limited network. 
	The plant is described by the following linear discrete-time dynamics
	\begin{subequations}\label{eq:dissys}
		\begin{align}
		x_{s+1} &= A x_{s} + B u_{s} \label{eq:dissys1}\\
		y_s &= Cx_{s} \label{eq:dissys2}
		\end{align}
	\end{subequations}
where $x_s \in \mathbb{R}^{n_x}, u_s \in \mathbb{R}^{n_u}$, and $y_s \in \mathbb{R}^{n_y}$ denote the state, control input, and output, respectively.
	Here, the plant is connected to a sensor that samples the output $y_s$ at time instants $\{s_\ell\}_{\ell \in \mathbb{N}_{0}}\subseteq \{s\}_{s \in \mathbb{N}_{0}}$ dictated by a self-triggering mechanism which we design later. 
The sampled output $y_{s_\ell}$ then passes through an encoder and gets quantized before sent to the self-triggering module and the controller.
	Using the quantized output, the self-triggering module then calculates the next sampling time and sends it back to the encoder side.
	
	We consider in this paper that the C-A channel is ideal with sufficient bandwidth and low communication cost.
	In other words, neither quantized nor self-triggered communication is required for the C-A channel.
	We adopt a standard observer-based state feedback controller that receives quantized output measurements from the network, and generates as well as transmits control inputs to a zero-order holder (ZOH) before entering the plant.

	Before moving on, we make the following standard assumptions on the system \eqref{eq:dissys}.
	
	\begin{assumption}[\emph{Controllability and observability}]\label{as:abca}
		The pair $(A,B)$ is controllable, and the pair $(C,A)$ is observable.
	\end{assumption}

	\begin{assumption}[\emph{Initial state bound}]\label{as:initialx}
		An upper bound on the initial state $\Vert x_0\Vert_{\infty} \le E_{in}$ is known.
	\end{assumption}
	
	\begin{remark}
		In fact, there are a number of ways for obtaining upper bounds on the initial state, including, e.g., the zooming-out method in \cite{Liberzon2003On}.
	\end{remark}
	
	For clarity, this paper investigates the following closed-loop stability, as also used in \cite{Wakaiki2021selftrigger}.
	\begin{definition}\label{def:converge}
			The discrete-time system \eqref{eq:dissys} is exponentially stable under Assumptions \ref{as:abca}---\ref{as:initialx} if there exist constants $\Omega \ge 1$ and $\omega \in (0,1)$ independent of $E_{in}$, such that $\Vert x_{s}\Vert_{\infty} \le \Omega E_{in} \omega^{s}$ holds for all $s\in \mathbb{N}_{0}$ 
	\end{definition}
	In addition, we also call for the following norm introduced in \cite{Wakaiki2021selftrigger} for stability analysis of event-triggered control systems.
	\begin{lemma}[{\cite{Wakaiki2021selftrigger}},Lem. 3.2]
		\label{lem:norm}
		If there exists a matrix $G\! \in\! \mathbb{R}^{n_x \!\times n_x}$, constants $\Gamma > 1$, and $\gamma > 0$ such that $\Vert G^s\Vert_{\infty} \le \Gamma \gamma^s$ holds for all $s \in \mathbb{N}_{0}$,
		then the function $\Vert \cdot \Vert_{G}  : \mathbb{R}^{n_x} \rightarrow \mathbb{R}$ defined by $x \mapsto \Vert x\Vert_{G} := \sup_{s \in \mathbb{N}_{0}} \Vert \gamma^{-s} G^s x\Vert_{\infty}$
		is a norm on $\mathbb{R}^{n_x}$.
		Moreover, for every $x\in \mathbb{R}^{n_x}$ and $s \in \mathbb{N}_{0}$, it holds that $\Vert x\Vert_{\infty} \le \Vert x\Vert_G \le \Gamma\Vert x\Vert_{\infty},~{\rm and}~
		\Vert G^s x\Vert_{G}\le \gamma^s \Vert x\Vert_{G}$.
	\end{lemma}

	\section{Self-triggered Control: DoS-free Case}\label{sec:abdos}
	For stabilization of linear systems with output available only, it is common to adopt an observer-based feedback control law, namely using an observer to estimate the state and then using the estimated state to construct a state feedback controller.
	In addition, to compensate for the coarse transmission caused by the limited bandwidth, a dynamic quantizer is equipped such that the real output is quantized before transmitted.
	In this section, we begin by developing a self-triggered observer-based state feedback controller for linear systems with quantized output.
	When a faster transmission rate can be afforded at the C-A channel, e.g., $b \in \mathbb{N}_0$ time steps at the S-C channel equal one time step at the S-C channel, an observer-based deadbeat controller is proposed to simplify the self-triggering mechanism while further saving the communication resources.
	
	\subsection{Standard Observer-based Control}\label{sec:lq}
	
	The commonly used standard observer-based  controller (see e.g., \cite{OreillyJ}) is described as follows
	\begin{subequations}\label{eq:lqctrl}
		\begin{align}
		\hat{x}_{s+1}& =  A\hat{x}_{s} + B u_{s} +L[\mathcal{Q}(y_{s_\ell}) - \hat{y}_{s}], \label{eq:lqctrl1}\\
		\hat{y}_{s} &= C \hat{x}_{s} \label{eq:lqctrl2}\\
		u_{s} &= K \hat{x}_{s} \label{eq:lqctrl3}
		\end{align}
	\end{subequations}
	with the initial condition $\hat{x}_0 = 0$, where $\mathcal{Q}(y_{s_\ell})$ denotes the quantized output received from the network at previous self-triggered time $s_\ell$ (which is the most recent measurement at the controller side).
	The gain matrices $L$ and $K$ are determined such that $A - LC$ and $A + BK$ are Schur stable.
	Equations \eqref{eq:lqctrl1} and \eqref{eq:lqctrl2} constitute the standard Luenberger observer, and \eqref{eq:lqctrl3} follows a feedback control law based on the estimated state $\hat{x}_{s}$.
	The rest of this subsection is devoted to designing a self-triggering mechanism as well as  associated encoding scheme for a dynamic quantizer. 
	
	Let $e^{x}_{s} := x_{s} - \hat{x}_{s}$ denote the error between the estimated state and the actual state.
	Suppose we can construct a sequence $\{E^{x}_{s_\ell} \ge 0\}_{\ell\in \mathbb{N}_{0}}$ satisfying
	\begin{equation}\label{eq:lqeE}
	\Vert e^{x}_{s_\ell}\Vert_{\infty} \le E^{x}_{s_\ell},\quad \forall \ell\in\mathbb{N}_{0}.
	\end{equation}
	It follows from the definition of the induced $\infty$-norm that 
	\begin{equation}\label{eq:lqyE}
	\Vert y_{s_\ell} - \hat{y}_{s_\ell}\Vert_{\infty}  = \Vert C e^{x}_{s_\ell} \Vert_{\infty}\le \Vert C\Vert_{\infty} E^{x}_{s_\ell} \triangleq E_{s_\ell} .
	\end{equation}

	Consider the quantizer has e.g., $N$ quantization levels. 
	We take the quantization range to be $E_{s_\ell}$, and the quantization center to be $\hat{y}_{s_\ell}$ that can be provided by running in parallel \eqref{eq:lqctrl} at the encoder side. The dynamic output quantizer is presented as follows.
	
	At each self-triggering time $s_\ell$, the encoder partitions the hypercube $\{y \in \mathbb{R}^{n_y} : \Vert y_{s_\ell} - \hat{y}_{s_\ell}\Vert_{\infty} \le E_{s_\ell}\}$
	into $N^{n_y}$ equal-sized boxes, and uses a value in $\{1,2, \cdots\!,$ $N^{n_y} \}$ to one-to-one index each of those boxes.
	Let $\mathcal{Q}_{s_\ell}^{\rm ind}$ denote the partitioned box(es) containing $y_{s_\ell}$, which is sent to the decoder.
	If $y_{s_\ell}$ is on the boundary of multiple boxes, then any of the corresponding indices of these boxes can be used.
	The decoder receives the index $\mathcal{Q}_{s_\ell}^{\rm ind}$, and can easily recover the quantized output $\mathcal{Q}(y_{s_\ell})$ thanks to the one-to-one correspondence.
	Therefore, the quantization error of $y_{s_\ell}$ can be bounded by
	\begin{equation}\label{eq:yerror}
	\Vert y_{s_\ell} - \mathcal{Q}(y_{s_\ell})\Vert_{\infty} \le \frac{\Vert C\Vert_{\infty}}{N}E^{x}_{s_\ell}  =\frac{E_{s_\ell}}{N}.
	\end{equation}
	
	\subsubsection{Self-triggering scheme}\label{sec:lqselftri}
	In this part, we introduce a self-triggering mechanism that determines the sampling times $\{s_\ell\}_{\ell \in \mathbb{N}_{0}}$ based on the quantized output $ \mathcal{Q}(y_{s_\ell})$.
	
	Similarly to the event-triggered state feedback control in \cite{heemels2013model}, a general self-triggering mechanism computes the inter-sampling time $\tau_{\ell} := s_{\ell + 1} - s_\ell$ by constructing a function measuring some sort of deviation between the current output $y_s$ and the most recently sampled output $y_{s_\ell}$. 
	However, due to the limited datarate, only the quantized output $\mathcal{Q}(y_{s_\ell})$ is available.
		Therefore, we consider the following self-triggering mechanism
	\begin{equation}\label{eq:lqselftri}
	\bigg\{
	\begin{aligned}
	&s_{\ell + 1}\! :=\! s_\ell + \min\{\tau_{\max}, \tau_{\ell}\},~~ \ell \in \mathbb{N}_{0},~~ s_0 := 0\\
	&\tau_{\ell} \!:=\! \min\!\big\{\tau\in \mathbb{N}: g(\mathcal{Q}(y_{s_\ell}), E_{s_\ell}, \hat{x}_{s_\ell}, \tau) \!>\! \sigma E_{s_\ell}\big\}
	\end{aligned}
	\end{equation}
	where $\sigma > 0$ is a threshold, and $\tau_{\max}$ is an upper bound on the inter-sampling times $s_{\ell + 1} - s_\ell$ for all $\ell \in \mathbb{N}_{0}$.
	The function $g$ measures the `informativeness' of the current output $y_s$ relative to the previous quantized output $\mathcal{Q}(y_{s_\ell})$. Obviously, if the value of function $g$ is large enough, a new triggering time occur, and $y_s$ will be quantized and sent to the controller. The next lemma provides a way to construct such a triggering function.
	\begin{lemma}\label{lem:lqselftrig}
		Let Assumptions \ref{as:abca}---\ref{as:initialx} hold.
		Consider system \eqref{eq:dissys} adopting the controller \eqref{eq:lqctrl} and a quantizer such that condition  \eqref{eq:lqyE} is satisfied with a sequence $\{E_{s_\ell}\}_{\ell \in \mathbb{N}_{0}}$ for all $\ell \in \mathbb{N}_0$.
		If the self-triggering function $g$ in \eqref{eq:lqselftri} is given by
\begin{align}\label{eq:lqg}
&g(q, E,\hat{x},\tau) := \Big(\frac{\Vert C(A^{\tau} - I)\Vert_{\infty}}{\Vert C\Vert_{\infty}} + \frac{1}{N}\Big)E \\
&+ \Big\Vert C\Big(\!(A^{\tau} - I)\!+\! \sum_{i = 0}^{\tau - 1}A^{i}BK(A \!+ \!BK \!-\! LC)^{\tau - i - 1}\Big)\hat{x} \nonumber\\
& +  C\sum_{i = 0}^{\tau - 1}A^i BK\sum_{j = 0}^{\tau - j - 2}(A\! +\! BK \!-\! LC)^{i}Lq\Big\Vert_{\infty} , \tau \in   \mathbb{N}\nonumber
\end{align}	
		then the output error $\Vert \mathcal{Q}(y_{s_\ell}) - y_s\Vert_{\infty}$ satisfies
		\begin{equation}\label{eq:lqtriinequ}
		\Vert \mathcal{Q}(y_{s_\ell}) - y_s \Vert_{\infty} \le \sigma E_{{s_\ell}}, \quad \forall s \in [s_\ell, s_{\ell + 1}).
		\end{equation}	
	\end{lemma}
	\begin{pf}
		We start the proof by deriving an upper bound on the output quantization error $\Vert \mathcal{Q}(y_{s_\ell}) - y_s \Vert_{\infty} $.
		It follows from \eqref{eq:lqctrl} that $\hat{x}_{s_\ell + 1} = (A + BK - LC)\hat{x}_{s_\ell} + L\mathcal{Q}(y_{s_\ell})$.
		 Let an integer $p$ satisfy $p \in [1, s_{\ell + 1} - s_\ell)$. It can be recursively deduced that 
		\begin{align}\label{eq:lqhatxqll}
		\hat{x}_{s_\ell + p} &= (A + BK - LC)^p \hat{x}_{s_\ell}\nonumber\\
		&\quad   + \sum_{i = 0}^{p - 1}(A + BK - LC)^i L\mathcal{Q}(y_{s_\ell}).
		\end{align}
		From \eqref{eq:dissys}, one gets that
		\begin{align}\label{eq:lqxqell}
		x_{s_\ell + p} = A^px_{s_\ell} + \sum_{i = 0}^{p - 1}A^i BK \hat{x}_{s_\ell + p - i - 1}.
		\end{align}
		Substituting \eqref{eq:lqhatxqll} into \eqref{eq:lqxqell}, we arrive at
		\begin{align}
		x_{s_\ell + p} & = A^{p}x_{s_\ell}\!+\! \sum_{i = 0}^{p - 1}A^i BK\big[(A \!+\! BK \!-\! LC)^{p - i - 1}\hat{x}_{s_\ell}\nonumber \\
		&\quad + \sum_{j = 0}^{p - i - 2}(A \!+\! BK \!-\! LC)^j L \mathcal{Q}(y_{s_\ell})\big]\nonumber\\
		& = A^p x_{s_\ell} \!+ \!\sum_{i = 0}^{p - 1}A^i BK (A \!+\! BK\! -\! LC)^{p - i - 1}\hat{x}_{s_\ell}\nonumber\\
		&\quad  + \!\!\sum_{i = 0}^{p - 1}\!\!A^iBK\!\!\!\sum_{j = 0}^{p - i - 2}\!\!\!(\!A \!+\! BK \!-\! LC)^j\! L\mathcal{Q}(y_{s_\ell}).\label{eq:lqxsellp}
		\end{align}
		Moreover, 
		\begin{align*}
		x_{s_\ell + p} - x_{s_\ell} &= (A^p - I)(x_{s_\ell} - \hat{x}_{s_\ell})+ \Big[(A^p - I) \\
		&\quad + \sum_{i = 0}^{p - 1}A^i BK (A + BK - LC)^{p - i - 1}\Big]\hat{x}_{s_\ell} \\
		&\quad+ \!\sum_{i = 0}^{p - 1}\!A^{i}BK\!\!\!\sum_{j = 0}^{p - i - 2}\!\!\!(A \!+\! BK \!- \!LC)^{j}L\mathcal{Q}(y_{s_\ell}).
		\end{align*}
		Therefore, the output error satisfies
		\begin{subequations}\label{eq:yerror2}
			\begin{align}
		&\quad~ \Vert y_{s_\ell + p} - \mathcal{Q}(y_{s_\ell})\Vert_{\infty} \label{eq:yerror2_1}\\
		&  = \Vert C(x_{s_\ell + p} - x_{s_\ell}) + y_{s_\ell} - \mathcal{Q}(y_{s_\ell})\Vert_{\infty}\label{eq:yerror2_2}\\
		& \le \Big(\frac{\Vert C(A^{p} - I)\Vert_{\infty}}{\Vert C\Vert_{\infty}} + \frac{1}{N}\Big)E_{s_\ell} \label{eq:yerror2_3}\\
		&\quad+ \Big\Vert C\Big(\!A^{p} \!-\! I \!+ \!\sum_{i = 0}^{p - 1}\!A^{i}BK(A \!+\! BK\! -\! LC)^{p - i - 1}\Big)\hat{x}_{s_\ell} \nonumber\\
		&\quad + C\sum_{i = 0}^{p - 1}A^i BK\!\!\sum_{j = 0}^{p - j - 2}\!\!(A \!+ BK \!- LC)^{i}L\mathcal{Q}(y_{s_\ell})\Big\Vert_{\infty}\nonumber
		\end{align}
		\end{subequations}
	where \eqref{eq:yerror2_3} is derived by substituting \eqref{eq:lqyE} and \eqref{eq:yerror} into \eqref{eq:yerror2_2}.
	Furthermore, according to  \eqref{eq:lqg}, it can be deduced that $\Vert \mathcal{Q}(y_{s_\ell}) - y_s \Vert_{\infty} \le g(\mathcal{Q}(y_{s_\ell}), E_{s_\ell}, \hat{x}_{s_\ell}, p)$.
		Finally, noticing from \eqref{eq:lqselftri} that $g(\mathcal{Q}(y_{s_\ell}), E_{s_\ell}, \hat{x}_{s_\ell}, p) \le \sigma E_{s_\ell}$, we have that $\Vert \mathcal{Q}(y_{s_\ell}) - y_s \Vert_{\infty} \le g(\mathcal{Q}(y_{s_\ell}), E_{s_\ell}, \hat{x}_{s_\ell}, p) \le \sigma E_{s_\ell}$, thus completing the proof.
	\end{pf}
	\subsubsection{Encoding scheme}
	\label{sec:lqencoding}
	When adopting a dynamic quantizer, the quantization center $\hat{y}_{s_\ell}$ and the range $E_{s_\ell}$ should evolve with time to ensure that the hypercube always contains the actual output $y_{s_\ell}$.
	Recall that the center $\hat{y}_{s_{\ell}}$ is obtained by running the controller \eqref{eq:lqctrl} at the encoder side in parallel.
	We here design an update rule for the sequence $\{E_{s_\ell}\}_{\ell \in \mathbb{N}_{0}}$ such that \eqref{eq:lqyE} is satisfied.
	
	Since matrix $L$ is chosen such that $A_{cl} := A - LC$ is Schur stable, there exist constants $\Gamma \ge 1$ and $\gamma \in (0,1)$ such that 
	\begin{align}\label{eq:gammaGamma}
	\Vert (A - LC)^s\Vert_{\infty} \le \Gamma \gamma^{s},\quad \forall s \in \mathbb{N}_{0}.
	\end{align}
	Choosing $\alpha := {\Gamma\Vert L\Vert_{\infty}\Vert C\Vert_{\infty}}/(1 - \gamma)$, 
	we define $\{E_{s_\ell}\}_{\ell \in \mathbb{N}_{0}}$ as follows
	\begin{equation}\label{eq:lqE}
	\begin{cases}
	E_{s_\ell} := \Vert C\Vert_{\infty} E^{x}_{s_\ell}\\
	E^{x}_{0} := \Gamma E_{in}\\
	E^{x}_{s_{\ell + 1}} \!:=\! (\gamma^{s_{\ell + 1} \!-\! s_\ell}(1 - \alpha \sigma) \!+\! \alpha \sigma)E^{x}_{s_\ell}, ~\forall\ell\! \in\!  \mathbb{N}_{0} 
	\end{cases}
	\end{equation}
	where $\sigma$ is the threshold parameter in \eqref{eq:lqselftri}.
	
	The following result proves that the output quantization error is upper-bounded by the sequence \eqref{eq:lqE}, and it converges to the origin exponentially fast.
	
	\begin{lemma}\label{lem:lqbound}
		Let Assumptions \ref{as:abca}---\ref{as:initialx} hold.
		Consider system \eqref{eq:dissys} with the standard controller \eqref{eq:lqctrl}. 
		Suppose that the sampling times $\{s_\ell\}_{\ell \in \mathbb{N}_{0}}$ are generated by \eqref{eq:lqselftri} with i) the self-triggering function $g$ in \eqref{eq:lqg}, ii) the sequence $\{E_{s_\ell}\}_{\ell \in \mathbb{N}_{0}}$ in \eqref{eq:lqE}, and iii) the threshold $\sigma > 0$ satisfying
		\begin{equation}\label{eq:sigmacondition}
		\frac{1}{N} \le \sigma < \frac{1}{\alpha} = \frac{1 - \gamma}{\Gamma \Vert L\Vert_{\infty}\Vert C\Vert_{\infty}}.
		\end{equation}
		Then, the following statements hold:
		\begin{itemize}
			\item [s1)]
			$\Vert x_{s_\ell + p} - \hat{x}_{s_\ell + p}\Vert_{A_{cl}} \le (\gamma^{p}(1 - \alpha \sigma) + \alpha \sigma)E^{x}_{s_\ell}$, where $p \in [0, s_{\ell + 1} - s_{\ell}]$; 
		\item [s2)]
				For $\tau \in \mathbb{R}_{\ge 0}$, function $\left(\gamma^{\tau}(1-\alpha \sigma)+\alpha \sigma\right)^{1 / \tau}$ is strictly increasing, and 
			$\tau_{\max }=\underset{1 \leq \tau \leq \tau_{\max }}{\operatorname{argmax}}\left(\gamma^{\tau}(1-\delta \sigma)+\delta \sigma\right)^{1 / \tau}$; 
			\item [s3)]
			$\Vert y_{s_\ell} - \hat{y}_{s_\ell}\Vert_{\infty}\le E_{s_{\ell}}
		\le \Gamma_1 E_{in} \omega_1^{s_{\ell}}$ where $\ell \in \mathbb{N}_{0}$, $\omega_1 :=  (\gamma^{\tau_{\max}}(1 - \alpha\sigma)+ \alpha\sigma)^{1/\tau_{\max}} < 1$, and $\Gamma_1 := \Vert C\Vert_{\infty}\Gamma$. 
		In particular, the sequence $\{E_{s_\ell}\}_{\ell\in \mathbb{N}_{0}}$ converges exponentially to the origin. 
		\end{itemize}
	\end{lemma}
\begin{pf}
	We begin by proving statement (s1).
	Based on the norm defined in Lem. \ref{lem:norm}, $e^{x}_{s_\ell} = x_{s_\ell} - \hat{x}_{s_\ell}$ satisfies
	\begin{align}\label{eq:normproperty1}
	\Vert e^{x}_{s_\ell}\Vert_{\infty} \le \Vert e^{x}_{s_\ell}\Vert_{A_{cl}} \le \Gamma \Vert e^{x}_{s_\ell} \Vert_{\infty} 
	\end{align}
	and
	\begin{equation}\label{eq:normproperty2}
	\Vert A_{cl}^q e^{x}_{s_\ell}\Vert_{A_{cl}} \le \gamma^q\Vert e^{x}_{s_\ell} \Vert_{A_{cl}},~\forall q \in \mathbb{N}_{0}.
	\end{equation}
	According to Assumption \ref{as:initialx} and \eqref{eq:lqE}, we have that $\Vert x_0\Vert_{A_{cl}} \le \Gamma \Vert x_0\Vert_{\infty} = E^{x}_{0}$.
	Supposing that $\Vert e^{x}_{s_\ell}\Vert_{A_{cl}} \le E^{x}_{\ell}$ holds for $s_\ell$, we use induction to prove that it holds for  $s_{\ell + 1}$.
	It follows from \eqref{eq:dissys} and \eqref{eq:lqctrl} that $x_{s_\ell + 1} - \hat{x}_{s_\ell + 1} = A_{cl}(x_{s_\ell} - \hat{x}_{s_\ell}) - L(\mathcal{Q}(y_{s_\ell}) - \hat{y}_{s_\ell})$.
	Let $p \in [0, s_{\ell + 1} - s_\ell]$.
	Recursively, it can be deduced that
	\begin{subequations}\label{eq:eacl}
		\begin{align}
		&\quad~ \Vert x_{s_\ell + p} - \hat{x}_{s_\ell + p}\Vert_{A_{cl}}\nonumber\\
		& = \Big\Vert A_{cl}^p (x_{s_\ell} \!-\! \hat{x}_{s_\ell}) \!-\! \sum_{i = 0}^{p - 1}A_{cl}^i L (\mathcal{Q}(y_{s_\ell}) \!-\! y_{s_\ell + p - i - 1})\Big\Vert_{A_{cl}}\nonumber\\
		& \le \gamma^p \Vert x_{s_\ell} - \hat{x}_{s_\ell}\Vert_{A_{cl}} \nonumber\\
		&~~~ + \Gamma \sum_{i = 0}^{p - 1}\Vert A_{cl}^i L\Vert_{\infty}\Vert \mathcal{Q}(y_{s_\ell}) - y_{s_\ell + p - i - 1}\Vert_{\infty}\label{eq:eacl1}\\
		&\le \gamma^pE^{x}_{s_\ell} + \sigma \Gamma \Vert L\Vert_{\infty}\Vert C\Vert_{\infty}\sum_{i = 0}^{p - 1}\gamma^{i} E^{x}_{s_\ell}\label{eq:eacl2}\\
		& \le (\gamma^{p}(1 - \alpha \sigma) + \alpha \sigma)E^{x}_{s_\ell}\label{eq:eacl3}
		\end{align}
	\end{subequations}
	where inequality \eqref{eq:eacl1} follows from \eqref{eq:normproperty1} and \eqref{eq:normproperty2}.
	According to Lem. \ref{lem:lqselftrig}, we arrive at \eqref{eq:eacl2}.
	When $p = s_{\ell + 1} - s_{\ell}$, inequality $\Vert x_{s_{\ell + 1}} - \hat{x}_{s_{\ell + 1}}\Vert_{A_{cl}} \le E^{x}_{s_{\ell + 1}}$ arises by substituting \eqref{eq:lqE} into \eqref{eq:eacl3}.
	Therefore, $\Vert e^{x}_{s_\ell}\Vert_{A_{cl}} \le E^{x}_{s_\ell}$ holds for all $\ell \in \mathbb{N}_{0}$.
	Noticing that $\Vert e^{x}_{s_\ell}\Vert_{\infty} \le \Vert e^{x}_{s_\ell}\Vert_{A_{cl}}$, hence $\Vert e^{x}_{s_\ell}\Vert_{\infty} \le E^{x}_{s_\ell}$ holds for all $\ell \in \mathbb{N}_{0}$.
	
Statement (s2) can be derived from the proof of \cite[Thm. 3.5]{Wakaiki2021selftrigger} and is thus omitted here.

Next, since $\Vert y_{s_\ell} - \hat{y}_{s_\ell}\Vert_{\infty} = \Vert C(x_{s_\ell} - \hat{x}_{s_\ell})\Vert_{\infty} = \Vert Ce^{x}_{s_\ell}\Vert_{\infty}$ and $E_{s_\ell} = \Vert C\Vert_{\infty} E^{x}_{s_\ell}$, it can be deduced that $\Vert y_{s_\ell} - \hat{y}_{s_\ell}\Vert_{\infty} \le E_{s_\ell}$ is met.
In addition, although $\tau \in \mathbb{N}$ in \eqref{eq:lqselftri}, the self-triggering function $g$ should always satisfy $g(\mathcal{Q}(y_{s_\ell}), E_{s_\ell}, \hat{x}_{s_\ell}, 0) = (1/N)E_{s_\ell}\le \sigma E_{s_\ell}$; otherwise,  $g(\mathcal{Q}(y_{s_\ell}), E_{s_\ell}, \hat{x}_{s_\ell}, \tau) \le \sigma E_{s_\ell}$ does not hold even for $\tau = 1$.
Therefore, parameter $\sigma$ is chosen such that $\sigma \ge 1/N$.
Based on \eqref{eq:lqE}, it follows that
{\setlength\abovedisplayskip{3pt}	
	\setlength\belowdisplayskip{3pt}
	\begin{align}
E_{s_{\ell + 1}} & = \Vert C\Vert_{\infty}\Gamma E_{in} \prod_{i = 0}^{\ell}[\gamma^{s_{i + 1} - s_{i}}(1 - \alpha\sigma) + \alpha\sigma]\nonumber\\
& \le \Gamma_1 E_{in} \omega_1^{s_{\ell + 1}}\label{eq:omega1}
\end{align} }
where $\omega_1 := (\gamma^{\tau_{\max}}(1 - \alpha\sigma)+ \alpha\sigma)^{1/\tau_{\max}}$ and $\Gamma_1 := \Vert C\Vert_{\infty}\Gamma$.
It follows from \eqref{eq:sigmacondition} that $\alpha\sigma < 1$, and since $\gamma \in (0, 1)$, it can be deduced that $\gamma^{\tau}(1 - \alpha\sigma) + \alpha\sigma < \gamma^{0}(1 - \alpha\sigma) + \alpha\sigma = 1$
Therefore, $\omega_1 < 1$, and it follows from \eqref{eq:omega1} that sequence $\{E_{s_\ell}\}_{\ell \in \mathbb{N}_{0}}$ converges to the origin, which completes the proof.	
\end{pf}
	Based on Lems. \ref{lem:lqselftrig} and \ref{lem:lqbound}, the stability condition is now ready to be presented.
	\begin{theorem}\label{thm:lqconverge}
		Consider system \eqref{eq:dissys} with the  controller \eqref{eq:lqctrl}, where the matrices $L$ and $K$ are chosen such that $A - LC$ and $A + B K$ are Schur stable. Under Assumptions \ref{as:abca}---\ref{as:initialx}, if the sampling times $\{s_\ell\}_{\ell \in \mathbb{N}_{0}}$ are generated by \eqref{eq:lqselftri} with i) the self-triggering function $g$ in \eqref{eq:lqg}, ii) the sequence $\{E_{s_\ell}\}_{\ell \in \mathbb{N}_{0}}$ in \eqref{eq:lqE}, and iii) the threshold parameter $\sigma > 0$ satisfying \eqref{eq:sigmacondition}, then the system \eqref{eq:dissys} is exponentially stable.
		In particular, the decay parameter $\omega  \in (\hat{\omega}, 1)$ satisfies Def. \ref{def:converge} for some constant $\Omega > 0$ where $\hat{\omega } >\max \{\omega_1, \omega_2 \}$ with $\omega_2 \in (0,1)$ obeying $\Vert (A + B K)^s\Vert_\infty \ge \Omega_K \omega_2^s$ for some constant $\Omega_K >0$ and all $s \in \mathbb{N}_0$.
	\end{theorem}
\begin{pf}
	It follows from \eqref{eq:dissys1} that
	\begin{subequations}\label{eq:lqnormx}
		\begin{align}
			\Vert x_{s_\ell}\Vert_{\infty} &= \Big\Vert (A \!+\! B K)^{s_\ell}x_{0}\label{eq:lqnormx1}\\
			& ~~~ -\sum_{i = 0}^{s_\ell - 1}(A \!+\! B K)^{s_\ell - i - 1}B K e^{x}_{ i}\Big\Vert_{\infty}\nonumber\\
			& \le \Vert (A \!+\! B K)^{s_\ell}\Vert_{\infty}\Vert x_{0}\Vert_{\infty} \label{eq:lqnormx2}\\
			&~~~ +  \sum_{i = 0}^{s_\ell - 1}\Vert(A \!+\! B K)^{s_\ell - i - 1}B K\Vert_{\infty} \Vert e^{x}_{ i}\Vert_{\infty}\nonumber.
		\end{align}	
	\end{subequations}	
	
	\noindent	Since $A + B K$ is Schur stable, there exist constants $\Omega_K >0$, and $\omega_2 \in (0,1)$ such that $\Vert (A + B K)^s\Vert_{\infty} \le \Omega_K \omega_2^s$ for all $s \in \mathbb{N}_{0}$.
	In addition, noticing that $\gamma \in (0,1)$, it can be deduced from the statement (s1) of Lem. \ref{lem:lqbound} that for all $\ell \in \mathbb{N}_0$, and $p \in [0,s_{\ell+1}-s_{\ell})$ 
	\[
	\|e^x_{s_{\ell}+p}\|_{\infty} \leq E^x_{s_{\ell}}
	\leq \omega_1^{-\tau_{\max}} \Gamma E_{in}\omega_1^{s_{\ell}+p}.
	\]
	Hence,
	\begin{subequations}\label{eq:lqxconverge}
		\begin{align}
			&\sum_{i=0}^{s_{\ell}-1} 
			\|(A+BK)^{s_{\ell}-i-1} BK\|_{\infty} \|e^x_i\|_{\infty}\nonumber\\
			&\leq
			\sum_{i=0}^{s_{\ell}-1}\! 
			\big(
			\|BK\|_{\infty} 
			\Omega_K\omega_2^{s_{\ell}-i-1} 
			\big)\! \cdot\! \big(
			\omega_1^{-\tau_{\max}} \Gamma E_{in}\omega_1^{i}
			\big) \label{eq:lqxconverge1}\\
			&\leq
			\big(
			\underbrace{\|BK\|_{\infty} 
				\Omega_K \omega_1^{-\tau_{\max}} \Gamma}_{\stackrel{\triangle}{=}\Omega_1} E_{in}
			\big)
			\sum_{i=0}^{s_{\ell}-1} \omega_2^{s_{\ell}-i-1} \omega_1^{i} \label{eq:lqxconverge2}\\
			&\leq
			\Omega_1 E_{in}
			\sum_{i=0}^{s_{\ell}-1} \omega^{s_{\ell}-1},\quad \omega :=\max\{\omega_1, \omega_2\}, \label{eq:lqxconverge3}\\
			&\leq 
			\Omega_1 E_{in}
			(s_\ell \omega^{s_{\ell}-1}).\label{eq:lqxconverge4}
		\end{align}	
	\end{subequations} 	
	Notice that for all $\hat \omega > \omega$, there exists a constant $c_1>0$
	such that
	$s\omega^s \leq c_1 \hat \omega^s$ for all $s \in \mathbb{N}_0$.
	Moreover, 
	\begin{equation}\label{eq:omega12}
		\Vert (A \!+\! B K)^{s_\ell}\Vert_{\infty}\Vert x_{0}\Vert_{\infty}\le \Omega_K E_{in}\omega^{s_{\ell}}
	\end{equation} 
	thus we have from \eqref{eq:lqnormx} that
	\begin{equation}\label{eq:lqxsell}
		\|x_{s_\ell}\|_{\infty} \leq \hat{\Omega}_1 E_{in} \hat{\omega}^{s_{\ell}}
	\end{equation}
	for some $\hat{\Omega}_1 >0$.
	
	Having shown the exponential convergence for $s = s_{\ell}$, we next extend \eqref{eq:lqxsell} to all $s\in \mathbb{N}_{0}$.
	It can be deduced from \eqref{eq:lqxsellp} that, for all $\ell \in \mathbb{N}_{0}$ and $p \in [0, s_{\ell + 1}-s_{\ell})$
	\begin{equation}
		\Vert x_{s_\ell + p}\Vert_{\infty} \le c_2\Vert x_{s_\ell}\Vert_{\infty} + c_3\Vert \hat{x}_{s_\ell}\Vert _{\infty} + c_4 \|\mathcal{Q}(y_{s_\ell})\|_{\infty}
	\end{equation}
	where constants $c_2, c_3, c_4 > 0$ are finite because $s_{\ell + 1} - s_{\ell} \le \tau_{\max}$.
	Moreover, since
	\begin{align*}
		\|\mathcal{Q}(y_{s_\ell})\|_{\infty} &\leq E_{s_\ell} + \|C\|_{\infty} \|\hat x_{s_\ell}\|_{\infty}\\
		\|\hat x_{s_\ell}\| &\leq \|x_{s_\ell}\|_{\infty} +  \|e^x_{s_\ell}\|_{\infty}
	\end{align*}
	there exists $c_5 >0$ such that 
	\begin{equation*}
		\|x_{s_{\ell}+p}\|_{\infty} \!\!\leq\! c_5 E_{in} \hat \omega^{s_\ell} \!=\!\!
		\big(c_5  \hat \omega^{\!-\!\tau_{\max}}\!
		\big)\!E_{in}
		\hat \omega^{s_\ell+p} \!=\! \hat{\Omega}_2 E_{in} \hat \omega^{s_\ell + p} 
	\end{equation*}
	for all $\ell \!\in \!\mathbb{N}_0$ and $p \!\in\! [0,s_{\ell+1}\!-\!s_{\ell})$.
	Hence, based on Def. \ref{def:converge}, system \eqref{eq:dissys} converges exponentially to the origin.
\end{pf}
	\begin{remark}[\emph{Computational resource}]
		Although \eqref{eq:lqctrl} achieves exponential stabilization for the self-triggered control system \eqref{eq:dissys} using quantized output, the quantizer requires running the controller \eqref{eq:lqctrl} at the encoder side.
		In addition, the self-triggering module using the estimated state $\hat{x}$ also needs to run the controller \eqref{eq:lqctrl}.
		This certainly challenges the practical implementation and increases the computational burden. As a remedy, one can simply choose the origin to be the quantization center.
		In that case, it has been shown that computational savings can be achieved by sacrificing the convergence rate in \cite{8880482}.
	Specifically, the origin quantization center may decrease the accuracy of the quantized data under the same data rate.
	In addition, coarser quantization may lead to smaller inter-sampling times, and consequently increase the communication overhead.
		To balance between the computational and communication loads as well as maintain a fast convergence rate, a deadbeat control scheme is advocated in the ensuing section.
	\end{remark}

\subsection{Deadbeat Control}\label{sec:db}
In the previous section, the observer-based state feedback controller should be copied at both the encoder side and the self-triggering module, which inevitably increases the computational overload.
Moreover, compared with the self-triggering mechanism suggested for quantized state feedback control in \cite{Wakaiki2021selftrigger}, condition \eqref{eq:lqg} entails an additional term, namely the estimated state $\hat{x}$.
Involving $\hat{x}$ in the self-triggering condition appears natural, because only quantized output is available here.
Nonetheless, we next show that this information can be eliminated from the condition to facilitate implementation of the self-triggering mechanism, by means of a simple deadbeat control protocol design.

	We assume in this section that the C-A channel can afford a higher communication rate.
Specially, the discrete-time system in \eqref{eq:dissys} is replaced by
\begin{subequations}\label{eq:dissys_2}
	\begin{align}
	x_{s,k+1} &= \tilde{A}x_{s,k} + \tilde{B}u_{s,k} \label{eq:dissys_21}\\
	y_{s,k} &= Cx_{s,k} \label{eq:dissys_22}
	\end{align}
\end{subequations}
where $s \in \mathbb{N}_{0}$ denotes the time index of the S-C channel (S-C time step), and $k = 0,\,1,\, \cdots\!,\, \eta-1$ is the index of the time step at the C-A channel (C-A time step) during one S-C time step; see Fig. \ref{fig:timestep} for an illustration.
The integer $\eta$ is the controllability index of system \eqref{eq:dissys_2}, which can be obtained by computing the smallest integer $\eta$ such that
${\rm rank} \big[\tilde{B},\,\tilde{A}\tilde{B}, \cdots, \tilde{A}^{\eta} \tilde{B}\big] = n_x$.
The relationship between matrices $A$, $B$, and $\tilde{A}$, $\tilde{B}$ are $A = \tilde{A}^\eta$, and $B = \sum_{i = 0}^{\eta - 1}\tilde{A}^i \tilde{B}$.	
For simplicity, we use $x_s$ to denote $x_{s,0}$, and the same for $\hat{x}_s$, $\hat{y}_s$ and $y_s$.
In addition, assuming the lengths of one time step of the S-C channel and the C-A channel are $\Delta$ and $\delta$, respectively, we work with
{\setlength\abovedisplayskip{3pt}	
	\setlength\belowdisplayskip{3pt \begin{equation}\label{eq:delta}
		\delta = {\Delta}/{\eta}
		\end{equation}}
\begin{remark}
	In the sampling-data setting, the discrete-time system matrices are calculated using the continuous-time system matrices and the sampling period, i.e., $\Delta$ and $\delta$; see, e.g., Remark 2.4 in \cite{Wakaiki2021selftrigger}.
	Hence, it is easy to construct matrices $\tilde{A}$ and $\tilde{B}$ such that $A = \tilde{A}^\eta$ and $B = \sum_{i = 0}^{\eta - 1}\tilde{A}^i \tilde{B}$ hold true.
\end{remark}

\vspace{-0.2cm}
	\begin{figure}
	\centering
	\includegraphics[width=7cm]{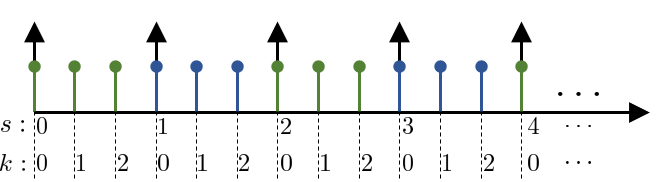}\\
	\caption{Relationship between the C-A time step and the S-C time step with $\eta = 3$.}\label{fig:timestep}
	\centering
\end{figure}

Similar to Assumption \ref{as:abca}, matrices $\tilde{A}$, $\tilde{B}$, and $C$ obey the following assumption.	
\begin{assumption}[\emph{Controllability and observability}]
	\label{as:delta_db}
	The pairs $(\tilde{A}, \tilde{B})$ and $(C, \tilde{A}^\eta)$ of the linear discrete-time system \eqref{eq:dissys_2} are controllable and observable.
\end{assumption}
The controller \eqref{eq:lqctrl} is now replaced by the observer-based deadbeat controller 
	\begin{subequations}\label{eq:dbctrl}
		\begin{align}
		&\hat{x}_{s, k+1} = \tilde{A}\hat{x}_{s,k} + \tilde{B}u_{s,k}, & k&\le \eta - 1 \label{eq:dbctrl1}\\
		&\hat{x}_{s} = \hat{x}_{s-1, \eta} + M[\mathcal{Q}(y_{s_{\ell}}) - \hat{y}_{s-1, \eta}], 
		\label{eq:dbctrl2}\\
		&\hat{y}_{s,k} = C \hat{x}_{s,k} \label{eq:dbctrl3}\\
		&u_{s,k} = K \hat{x}_{s,k} \label{eq:dbctrl4}
		\end{align}
	\end{subequations}
	with $\hat{x}_0 = 0$.
	The control input is generated by \eqref{eq:dbctrl4}.
Matrix $M \in \mathbb{R}^{n_x \times n_y}$ can be seen as the observer gain such that $\bar{A}_{cl} := \tilde{A}^{\eta}(I - MC)$ is Schur stable, which is always feasible for observable systems $(C, \tilde{A}^{\eta})$.
In particular, one can choose suitable $\bar{M} := \tilde{A}^{\eta}M$ such that $\tilde{A}^{\eta} - \bar{M}C$ is Schur stable, and $M = (\tilde{A}^\eta)^{-1} \bar{M}$.
In addition, since $(\tilde{A}, \tilde{B})$ is controllable, the controller gain $K \in \mathbb{R}^{n_u \times n_x}$ can be designed such that
{\setlength\abovedisplayskip{3pt}	
	\setlength\belowdisplayskip{3pt}
\begin{equation}\label{eq:dbdb}
(\tilde{A} + \tilde{B}K)^{\eta} = 0.
\end{equation}}
Let $\Vert e^{x}_{s -  1, \eta}\Vert_{\infty} := \Vert x_{s - 1, \eta} - \hat{x}_{s - 1, \eta}\Vert_{\infty}$, and construct the sequence $\{E_{s_\ell}\}_{\ell \in \mathbb{N}_{0}}$ obeying $E_{s_\ell} \ge \Vert C\Vert_{\infty} \Vert e^{x}_{s_\ell -  1, \eta}\Vert_{\infty}$ for all $\ell \in \mathbb{N}_{0}$.
Adopt the same quantizer described in Sec. \ref{sec:lq} with the quantization level $N$, the center $\hat{y}_{s_\ell - 1, \eta}$, and the range $E_{s_\ell}$.
It can be similarly shown that the quantization error $\Vert y_{s_\ell - 1, \eta} - \mathcal{Q}(y_{s_\ell - 1, \eta})\Vert_{\infty}$ satisfies \eqref{eq:yerror}.
Noticing from \eqref{eq:dbctrl1} that $\hat{x}_{s,\eta} = (\tilde{A} + \tilde{B} K)\hat{x}_{s,\eta - 1} = (\tilde{A} + \tilde{B} K)^{\eta}\hat{x}_{s} = 0$, 
one gets that $\hat{y}_{s,\eta} = C\hat{x}_{s,\eta} = 0$ for all $s \in \mathbb{N}_{0}$.
Therefore, it is sufficient to choose zero as the
quantization center, and the observer-based deadbeat controller \eqref{eq:dbctrl} does not need to be present at the encoder side. 
The updated self-triggering mechanism and associated output encoding scheme are presented as follows.
We consider the following self-triggering mechanism
\begin{equation}\label{eq:dbselftri}
\bigg\{
\begin{aligned}
&s_{\ell + 1} := s_\ell + \min\{\tau_{\max}, \tau_{\ell}\},~~ \ell \in \mathbb{N}_{0},~~ s_0 := 0\\
&\tau_{\ell} := \min\{\tau\in \mathbb{N}: \bar{g}(\mathcal{Q}(y_{s_\ell}), E_{s_\ell}, \tau) > \sigma E_{s_\ell}\}.
\end{aligned}
\end{equation}
In contrast to the self-triggering function $g$ in \eqref{eq:lqselftri}, function $\bar{g}$ here does not require the quantization center $\hat{x}_{s_{\ell}}$, but only the quantized output $\mathcal{Q}(y_{s_\ell})$ and the quantization range $E_{s_\ell}$.
This simplifies implementation and saves computational resources.
The following lemma presenting the self-triggering function $\bar{g}$ in \eqref{eq:dbselftri} is a counterpart of Lem. \ref{lem:lqselftrig}.
	\begin{lemma}\label{lem:dbselftrig}
		Let Assumptions \ref{as:initialx}, and \ref{as:delta_db} hold.
Consider system \eqref{eq:dissys} adopting the observer-based deadbeat controller in \eqref{eq:dbctrl}, and a quantizer such that condition \eqref{eq:lqyE} is satisfied with a sequence $\{E_{s_{\ell}}\}_{\ell \in \mathbb{N}_{0}}$ for all $\ell \in \mathbb{N}_{0}$.
	For every $s\in [s_{\ell}, s_{\ell + 1})$
	with $\ell \in \mathbb{N}_{0}$, 
	if the self-triggering function $\bar{g}$ in \eqref{eq:dbselftri} is chosen as
	\begin{subequations}\label{eq:dbg}
		\begin{align}
		\bar{g}(q,\!E,\!1)\! :=& \bigg[\!\frac{\Vert C\Phi_A(I\! -\! MC)\Vert_{\infty}}{\Vert C\Vert_{\infty}} \!+\! \frac{\Vert C\Phi_A M \Vert_{\infty}}{N} \!
		\bigg] E \!+\! \Vert q\Vert_{\infty} \\
		\bar{g}(q,\!E,\!\tau)\! := & \bigg[\!\frac{\Vert C\Phi_A\! \tilde{A}^{(\tau - 2)\eta}\!\bar{A}_{cl}\Vert_{\infty}}{\Vert C\Vert_{\infty}}\! +\! \frac{\Vert C\Phi_A\! \tilde{A}^{(\tau - 1)\eta}\!M\Vert_{\infty}}{N}\!\bigg]\!E \nonumber\\
		&+\Big\Vert \Big[C\Phi_A \sum_{i = 0}^{\tau - 2} \tilde{A}^{i\eta}M + I\Big]q\Big\Vert_{\infty}
		\end{align}
	\end{subequations}
where $\Phi_A = -\sum_{i = 0}^{\eta - 1}(\tilde{A} + \tilde{B}K)^i \tilde{B}K \tilde{A}^{\eta - i - 1}$ and $\tau \ge 2$,
	then the output error obeys
	\begin{equation}\label{eq:dbtriinequ}
\Vert \mathcal{Q}(y_{s_\ell}) - y_{s} \Vert_{\infty} \le \sigma E_{{s_\ell}}.
	\end{equation}	
\end{lemma}

Based on the self-triggering mechanism above, we construct a sequence $\{E_{s_\ell}\}$ to validate the condition in Lem. \ref{lem:dbselftrig}.
Select constants $\bar{\Gamma} \ge 1$ and $\bar{\gamma} \in (0,1)$ such that
$\Vert \bar{A}_{cl}^s\Vert_{\infty} =\Vert (\tilde{A}^\eta(I - MC))^s\Vert_{\infty} \le \bar{\Gamma}\bar{\gamma}^s$ for all $s \in \mathbb{N}_{0}$.
Update the sequence $\{E_{s_\ell}\}_{\ell \in \mathbb{N}_{0}}$ as follows
\begin{equation}\label{eq:dbE}
	\left\{
\begin{aligned}
&E_{s_\ell} := \Vert C\Vert_{\infty} E^{x}_{s_\ell}\\
&E^{x}_{0} := \bar{\Gamma} E^{x}_{in}\\
&E^{x}_{s_{\ell + 1}} = (\bar{\gamma}^{s_{\ell + 1} - s_\ell}(1 - \bar{\alpha} \sigma) + \bar{\alpha} \sigma)E^{x}_{s_\ell}, ~\ell \in \mathbb{N}_{0} 
\end{aligned}
\right.
\end{equation}
where $\bar{\alpha} := (\bar{\Gamma}\Vert \tilde{A}^\eta M\Vert_{\infty}\Vert C\Vert_{\infty})/(1 - \bar{\gamma})$.
The next result generalizes Thm. \ref{thm:lqconverge} by adopting the deadbeat controller in \eqref{eq:dbctrl}.

\begin{pf}
	Following the proof of Lem. \ref{lem:lqselftrig}, we give an upper bound on the output error $\Vert \mathcal{Q}(y_{s_\ell})  - y_s\Vert_{\infty}$.
	Substituting \eqref{eq:dbctrl1} into \eqref{eq:dissys_2}, one gets that
	\begin{align*}
		x_{s_\ell,k + 1} &= \tilde{A} x_{s_\ell, k} + \tilde{B} K \hat{x}_{s_\ell, k} \\
		&= (\tilde{A} + \tilde{B}K)x_{s_\ell,k} - \tilde{B}K(x_{s_\ell,k} - \hat{x}_{s_\ell, k})
	\end{align*} 
	and hence
	\begin{align*}
		x_{s_\ell, \eta} =~& (\tilde{A} + \tilde{B}K)^{\eta} x_{s_\ell}\\
		&- \sum_{i = 0}^{\eta - 1}(\tilde{A} + \tilde{B}K)^i\tilde{B}K(x_{s_\ell, \eta - i - 1} - \hat{x}_{s_\ell, \eta - i - 1})\\
		= & - \sum_{i = 0}^{\eta - 1}(\tilde{A} + \tilde{B}K)^i \tilde{B}K (x_{s_\ell, \eta - i - 1} - \hat{x}_{s_\ell, \eta - i - 1})
	\end{align*}
	where the second equation holds because $(\tilde{A} + \tilde{B}K)^{\eta} = 0$.
	Recursively, it can be deduced that for $p \in [0,s_{\ell+1} \!- \!s_{\ell})$
	\begin{align}\label{eq:xqellp}
		&x_{s_\ell + p, \eta}=\\
		& \! -\!\!\sum_{i = 0}^{\eta - 1}(\tilde{A}\! +\! \tilde{B}K)^i\!\tilde{B}K(x_{s_\ell + p, \eta - i - 1}\!-\! \hat{x}_{s_\ell + p, \eta - i - 1})\nonumber
	\end{align}
	and we only need to calculate $x_{s_\ell + p, \eta - i - 1} - \hat{x}_{s_\ell + p, \eta - i - 1}$.
	
	(c1) For every $s \in [s_\ell, s_{\ell + 1})$ and $k \in [1,\eta - 1]$, it follows from \eqref{eq:dissys_2} and \eqref{eq:dbctrl1} that
	$\hat{x}_{s, k + 1} = \tilde{A} \hat{x}_{s, k} + \tilde{B}K\hat{x}_{s, k}$, and $x_{s, k + 1} = \tilde{A}x_{s, k} + \tilde{B}K\hat{x}_{s, k}$.
	Therefore, 
	\begin{equation}\label{eq:dbxxhat}
		x_{s, k + 1} - \hat{x}_{s, k + 1} = \tilde{A}^{k + 1}(x_{s} - \hat{x}_{s}).
	\end{equation}

	(c2) For $s = s_\ell$, it can be obtained from \eqref{eq:dbctrl2} that, 
	\begin{subequations}
		\begin{align*}
			x_{s_\ell} &= x_{s_\ell - 1, \eta}\\
			\hat{x}_{s_\ell} &= \hat{x}_{s_\ell - 1, \eta} + M[\mathcal{Q}(y_{s_\ell}) - \hat{y}_{s_\ell - 1, \eta}].
		\end{align*}
	\end{subequations}	
	Hence, one gets that
	\begin{align*}
		x_{s_\ell} \!- \!\hat{x}_{s_\ell}& = x_{s_\ell} \!- \!\hat{x}_{s_\ell - 1, \eta} \!-\! M\big[\mathcal{Q}(y_{s_\ell}) \!-\! y_{s_\ell} \!+\! y_{s_\ell} \!-\! \hat{y}_{s_\ell - 1, \eta}\big]\\
		& = (I \!-\! MC)(x_{s_\ell} \!-\! \hat{x}_{s_\ell - 1, \eta})\! -\! M\big[\mathcal{Q}(y_{s_\ell})\!-\! y_{s_\ell}\big].
	\end{align*}
	
	(c3) For $s > s_\ell$, it can be deduced that
	\begin{subequations}\label{eq:haty}
		\begin{align}
			x_{s_\ell + p} &= x_{s_\ell + p - 1, \eta}\\
			\hat{x}_{s_\ell + p} &= \hat{x}_{s_\ell + p - 1, \eta} + M \mathcal{Q}(y_{s_\ell})\label{eq:haty0}
		\end{align}
	\end{subequations}
	where \eqref{eq:haty0} holds because $\hat{y}_{s_\ell + p - 1, \eta} = C \hat{x}_{s_\ell + p - 1, \eta} = C (\tilde{A} + \tilde{B}K)^{\eta}\hat{x}_{s_\ell + p - 1} = 0$.
	Hence,
	\begin{equation*}
		x_{s_\ell + p} - \hat{x}_{s_\ell + p} = (x_{s_\ell + p - 1, \eta} - \hat{x}_{s_\ell + p - 1, \eta}) - M \mathcal{Q}(y_{s_\ell}).
	\end{equation*}
	
	Substituting \eqref{eq:dbxxhat} into \eqref{eq:xqellp} yields
	\begin{equation*}
		x_{s_\ell + p, \eta} = \Phi_{A} (x_{s_\ell + p} - \hat{x}_{s_\ell + p})
	\end{equation*}
	where $\Phi_A = -\sum_{i = 0}^{\eta - 1}(\tilde{A} + \tilde{B}K)^i \tilde{B}K \tilde{A}^{\eta - i - 1}$ is defined in Lem. \ref{lem:dbselftrig}.
	In addition, noticing from (c1) and (c3) that
	\begin{align*}
		x_{s_\ell + p} - \hat{x}_{s_\ell + p} = \tilde{A}^{\eta}(x_{s_\ell + p - 1} - \hat{x}_{s_\ell + p - 1}) - M \mathcal{Q}(y_{s_\ell}).
	\end{align*}
	Combining this with (c2), recursively one deduces that
	\begin{align}\label{eq:xerror}
		&x_{s_\ell + p} - \hat{x}_{s_\ell + p}\nonumber\\
		= &~ \tilde{A}^\eta\big[\tilde{A}^\eta(x_{s_\ell + p - 2} - \hat{x}_{s_\ell + p - 2}) - M\mathcal{Q}(y_{s_\ell})\big] - M\mathcal{Q}(y_{s_\ell}) \nonumber \\
		= &~ \tilde{A}^{(p - 1)\eta}\big[\bar{A}_{cl}(x_{s_\ell} - \hat{x}_{s_\ell - 1, \eta})\big]\nonumber\\
		& - \tilde{A}^{p\eta}M\big[\mathcal{Q}(y_{s_\ell}) - y_{s_\ell}\big]- \sum_{i = 0}^{p - 1}\tilde{A}^{i\eta}M \mathcal{Q}(y_{s_\ell}).
	\end{align}
	Substituting \eqref{eq:xerror} into \eqref{eq:xqellp}, one has  that
	\begin{align*}
		x_{s_\ell, \eta} = \Phi_A [(I \!-\! MC)(x_{s_\ell} \!-\! \hat{x}_{s_\ell - 1, \eta}) \!-\! M(\mathcal{Q}(y_{s_\ell}) \!-\! y_{s_{\ell}})]
	\end{align*}
	and for $p \ge 1$,
	\begin{align*}
		&x_{s_\ell + p, \eta}  = \Phi_A \tilde{A}^{(p - 1)\eta}\bar{A}_{cl}(x_{s_\ell} - \hat{x}_{s_\ell - 1, \eta})\\
		&- \Phi_A \tilde{A}^{p\eta}M(\mathcal{Q}(y_{s_\ell}) - y_{s_\ell}) \!-\! \Phi_A \sum_{i = 0}^{p - 1}\tilde{A}^{i\eta}M\mathcal{Q}(y_{s_\ell}).
	\end{align*}
	Hence, the output error obeys,
	\begin{align*}
		&\Vert y_{s_\ell, \eta} - \mathcal{Q}(y_{s_\ell}) \Vert_{\infty} \nonumber\\
		&= \Big\Vert C\Phi_A(I-MC) (x_{s_\ell} - \hat{x}_{s_\ell - 1, \eta})\nonumber\\
		&~~~- C\Phi_A M[\mathcal{Q}(y_{s_\ell}) - y_{s_\ell}] -  \mathcal{Q}(y_{s_\ell})\Big\Vert_{\infty}\nonumber\\
		& \le \bigg[\frac{\Vert C\Phi_A(I\!-\!MC)\Vert_{\infty}}{\Vert C\Vert_{\infty}}\! +\! \frac{\Vert C\Phi_AM \Vert_{\infty}}{N} 
		\bigg] E_{s_{\ell}} \! + \!\Vert \mathcal{Q}(y_{s_\ell})\Vert_{\infty}
	\end{align*}
	and for $p \ge 1$, 
	\begin{align}
		&\Vert y_{s_\ell + p, \eta} - \mathcal{Q}(y_{s_\ell}) \Vert_{\infty} \nonumber\\
		&= \Big\Vert C\Phi_A \tilde{A}^{(p - 1)\eta}\bar{A}_{cl}(x_{s_\ell} - \hat{x}_{s_\ell - 1, \eta})\nonumber\\
		&~~~- C\Phi_A \tilde{A}^{p\eta}M[\mathcal{Q}(y_{s_\ell}) - y_{s_\ell}] \nonumber\\
		&~~~- \Big[C\Phi_A \sum_{i = 0}^{p - 1} \tilde{A}^{i\eta}M + I\Big]\mathcal{Q}(y_{s_\ell})\Big\Vert_{\infty}\nonumber\\
		&\le \bigg[\frac{\Vert  C\Phi_A \tilde{A}^{(p - 1)\eta}\bar{A}_{cl}\Vert_{\infty}}{\Vert C\Vert_{\infty}} + \frac{\Vert C\Phi_A \tilde{A}^{p\eta}M\Vert_{\infty}}{N}\bigg]E_{s_\ell}\nonumber\\
		&~~~+\Big\Vert \Big[C\Phi_A \sum_{i = 0}^{p - 1} \tilde{A}^{i\eta}M + I\Big]\mathcal{Q}(y_{s_\ell})\Big\Vert _{\infty}\label{eq:dbyerror}.
	\end{align}
	Eventually, according to the inequality  $\bar{g}(\mathcal{Q}(y_{s_\ell}), E_{s_\ell}, p) \le \sigma E_{s_\ell}$ in \eqref{eq:dbselftri} and noticing \eqref{eq:dbyerror}, if $\bar{g}$ is defined as in \eqref{eq:dbg}, then $\Vert \mathcal{Q}(y_{s_\ell}) - y_s \Vert_{\infty} \le \bar{g}(\mathcal{Q}(y_{s_\ell}), E_{s_\ell}, p) \le \sigma E_{s_\ell}$ is obtained using \eqref{eq:dbtriinequ}.
	This completes the proof.
\end{pf}
\begin{theorem}\label{thm:dbconverge}
	Consider system \eqref{eq:dissys} with the deadbeat controller \eqref{eq:dbctrl}, where the gain matrices $M$ and $K$ are chosen such that $\tilde{A}^\eta(I - MC)$ is Schur stable and $(\tilde{A} + \tilde{B}K)^{\eta} = 0$. 
	Under Assumptions \ref{as:initialx} and \ref{as:delta_db}, if the sampling times $\{s_\ell\}_{\ell \in \mathbb{N}_{0}}$ are generated by \eqref{eq:dbselftri} with i) the self-triggering function $\bar{g}$ in \eqref{eq:dbg}, ii) the sequence $\{E_{s_\ell}\}_{\ell \in \mathbb{N}_{0}}$ in \eqref{eq:dbE}, and iii) the threshold $\sigma > 0$ obeying
	\begin{equation}\label{eq:dbsigma}
	\frac{1}{N} \le \sigma < \frac{1}{\bar{\alpha}} = \frac{1 - \bar{\gamma}}{\bar{\Gamma} \Vert \tilde{A}^\eta M\Vert_{\infty}\Vert C\Vert_{\infty}}
	\end{equation}
	then the system \eqref{eq:dissys} is exponentially stable.
	In particular, the decay parameter $\omega= \bar{\gamma}^\tau_{\max} (1 - \bar{\alpha}\sigma) + \bar{\alpha}\sigma \in (0, 1)$ satisfies Def. \ref{def:converge} for some constant $\Omega > 0$.
\end{theorem} 
\begin{pf}
	We begin by showing that $\Vert y_{s_{\ell+ 1} - 1, \eta} - \hat{y}_{s_{\ell+ 1} - 1,\eta}\Vert_{\infty} \le E_{s_\ell + 1}$ for $\ell \in \mathbb{N}_{0}$.
	
	It follows from \eqref{eq:dissys_2} and \eqref{eq:dbctrl} that
	\begin{align*}
		x_{s_\ell, \eta} \!-\! \hat{x}_{s_{\ell}, \eta} \!=\! \bar{A}_{cl}(x_{s_{\ell} - 1, \eta} \!-\! \hat{x}_{s_{\ell} - 1, \eta})\! -\! \tilde{A}^\eta M (\mathcal{Q}(y_{s_\ell}) \!-\! y_{s_\ell}).
	\end{align*}
	Assuming that $\Vert x_{s_{\ell} - 1, \eta} - \hat{x}_{s_{\ell} - 1, \eta}\Vert_{A_{cl}} \le E_{s_\ell}^x$, recursively, one gets for $p \in [0, s_{\ell + 1} - s_\ell)$ that
	\begin{subequations}\label{eq:eacl_bar}
		\begin{align}
			&~~~\Vert x_{s_\ell + p, \eta} - \hat{x}_{s_\ell + p, \eta}\Vert_{\bar{A}_{cl}} \\
			& = \Big\Vert\bar{A}_{cl}^{p + 1}(x_{s_{\ell} - 1, \eta} - \hat{x}_{s_{\ell} - 1, \eta}) \nonumber\\
			&~~~- \sum_{i = 0}^{p}\bar{A}_{cl}^{i}\tilde{A}^\eta M(\mathcal{Q}(y_{s_\ell})- y_{s_\ell + p - i})\Big\Vert_{\bar{A}_{cl}}\\
			& \le \bar{\gamma}^{p + 1} \Vert x_{s_{\ell} - 1, \eta} - \hat{x}_{s_{\ell} - 1, \eta}\Vert_{A_{cl}} \nonumber\\
			&~~~ + \bar{\Gamma} \sum_{i = 0}^{p}\Vert\bar{A}_{cl}^i \tilde{A}^\eta M\Vert_{\infty}\Vert \mathcal{Q}(y_{s_\ell}) - y_{s_\ell + p - i}\Vert_{\infty}\label{eq:eacl_bar1}\\
			&\le \gamma^{p+1}E^{x}_{s_\ell} + \sigma \bar{\Gamma} \Vert \tilde{A}^\eta M\Vert_{\infty}\Vert C\Vert_{\infty}\sum_{i = 0}^{p}\bar{\gamma}^{i} E^{x}_{s_\ell}\label{eq:eacl_bar2}\\
			& \le (\bar{\gamma}^{p+1}(1 - \bar{\alpha} \sigma) + \bar{\alpha} \sigma)E^{x}_{s_\ell}\label{eq:eacl_bar3}\\
			& \le \bar{\omega}_1^{p+1}E^{x}_{s_\ell}, \quad \bar{\omega}_1 := \bar{\gamma}^{\tau_{\max}}(1 - \bar{\alpha} \sigma) + \bar{\alpha} \sigma \label{eq:eacl_bar4}
		\end{align}
	\end{subequations}
	where $\bar{\alpha} := (\bar{\Gamma}\Vert \tilde{A}^\eta M\Vert_{\infty}\Vert C\Vert_{\infty})/(1 - \bar{\gamma})$ is defined in \eqref{eq:dbE}.
	Since $\sigma < 1/\bar{\alpha}$, and $\bar{\gamma} < 1$, we deduce that $\bar{\gamma}^{p+1}(1 - \bar{\alpha} \sigma) + \bar{\alpha} \sigma \le \bar{\gamma}^{0}(1 - \bar{\alpha} \sigma) + \bar{\alpha} \sigma = 1$.
	Inequality \eqref{eq:eacl_bar4} follows from the statement (s2) in Lem. \ref{lem:lqbound}.
	Therefore, letting $p = s_{\ell  + 1} - s_\ell - 1$ in \eqref{eq:eacl_bar4} it follows from \eqref{eq:dbE} that
	\begin{align}\label{eq:dbomega1}
		\Vert y_{s_{\ell + 1} - 1, \eta} - \hat{y}_{s_{\ell + 1} - 1,\eta}\Vert_{\infty} &\le E_{s_{\ell + 1}}.
	\end{align}
	This is a counterpart of the statement (s3) in Lem. \ref{lem:lqbound}.
	
	Next, leveraging \eqref{eq:dbomega1}, we are able to prove the convergence of the state.
	Noticing from \eqref{eq:dbdb} that $\hat{x}_{s, \eta} = 0$ for every $s \in \mathbb{N}_{0}$.
	Hence, for $p \in [0, s_{\ell + 1} - s_\ell)$, according to \eqref{eq:dissys_2} and \eqref{eq:eacl_bar4}, we arrive at
	\begin{align}
		\Vert x_{s_\ell + p}\Vert_{\infty} &= \Vert x_{s_\ell + p - 1, \eta} - \hat{x} _{s_\ell + p - 1, \eta}\Vert_{\infty}\nonumber\\
		& \le  \bar{\Gamma}E_{in}\bar{\omega}_1^{s_{\ell} + p}.\label{eq:dbxp}
	\end{align}
	In addition, for $k \in [0,\eta - 1]$, one has from Lem. \ref{lem:dbselftrig} that
	\begin{subequations}\label{eq:dbx}
		\begin{align}
			&\Vert x_{s_\ell + p, k + 1}\Vert_{\infty} \nonumber\\ 
			&=\Big\Vert (\tilde{A} + \tilde{B} K)^{k + 1}x_{s_\ell + p} \nonumber\\
			&~~~ -\!\! \sum_{i = 0}^{k} (\tilde{A}\! +\! \tilde{B} K)^i\tilde{B}K\tilde{A}^{k - i} (x_{s_{\ell} + p} \!-\! \hat{x}_{s_\ell + p})\Big\Vert_{\infty}\label{eq:dbx_1}\\
			& = \Big\Vert (\tilde{A} + \tilde{B} K)^{k + 1}x_{s_\ell + p} \!- \!\!\sum_{i = 0}^{k} (\tilde{A} + \tilde{B} K)^i\tilde{B}K\tilde{A}^{k - i} \nonumber\\
			&~~~\times \big[(I - MC)(x_{s_{\ell} + p - 1,\eta} - \hat{x}_{s_\ell + p - 1, \eta}) \nonumber\\
			&~~~- M(\mathcal{Q}(y_{s_\ell})- y_{s_\ell + p})\big]\Big\Vert_{\infty}\label{eq:dbx_2}\\
			&\le \bar{\Omega}_K\Vert x_{s_\ell + p}\Vert_{\infty} \!+\! \big(\bar{\Omega}_2\bar{\Gamma}\!+\! \bar{\Omega}_3 \sigma \bar{\Gamma}_1\bar{\omega}_{1}^{-\tau_{\max}}\big)E_{in}\bar{\omega}_{1}^{s_{\ell}+p}\label{eq:dbx_3}\\
			&\le \bar{\Omega} E_{in}\bar{\omega}_{1}^{s_{\ell}+p}\label{eq:dbx_4}
		\end{align}
	\end{subequations}
	where $\bar{\Gamma}_1 := \Vert C\Vert_{\infty}\bar{\Gamma}$, and
	\begin{subequations}
		\begin{align*}
			&\bar{\Omega}_K := \max_{1 \leq k \leq \eta}
			\Vert (\tilde{A} + \tilde{B} K)^k\Vert_{\infty}\\
			&\bar{\Omega}_2 := \max_{0 \leq k \leq \eta - 1}
			\Vert \sum_{i = 0}^{k} (\tilde{A} + \tilde{B} K)^i\tilde{B}K\tilde{A}^{k - i}(I - MC)\Vert_{\infty}\\
			&\bar{\Omega}_3 := \max_{0 \leq k \leq \eta - 1}\Vert \sum_{i = 0}^{k} (\tilde{A} + \tilde{B} K)^i\tilde{B}K\tilde{A}^{k - i}M\Vert_{\infty}\\
			&\bar{\Omega} := (\bar{\Omega}_K + \bar{\Omega}_2 + \bar{\Omega}_3 \sigma \Vert C\Vert_{\infty}\bar{\omega}_{1}^{-\tau_{\max}})\bar{\Gamma}.
		\end{align*}
	\end{subequations}
	Equation \eqref{eq:dbx_1} holds by adopting \eqref{eq:dbxxhat}.
	In addition, equation \eqref{eq:dbx_2} holds by using \eqref{eq:haty}.
	Finally, substituting \eqref{eq:dbxp} into \eqref{eq:dbx_3}, the system is exponentially stable according to Def. \ref{def:converge}, which completes the proof.
\end{pf}
\begin{remark}{\bf {\rm \bf (}\emph{Comparison between the controllers \eqref{eq:lqctrl} and \eqref{eq:dbctrl}}{\rm \bf )}.}
	\label{rmk:twocontroller}
Relative to the standard observer-based controller in \eqref{eq:lqctrl}, the deadbeat controller in \eqref{eq:dbctrl} offers at least two advantages.
	\begin{itemize}
		\item [i)]
		The self-triggering mechanism requires only the quantized output $\mathcal{Q}(y_{s_\ell})$ and the quantization range $E_{s_\ell}$ to compute the next sampling time $s_{\ell + 1}$, whose implementation in practice is as simple as that for quantized state feedback control studied in \cite{Wakaiki2021selftrigger}.
		\item [ii)]
		At every sampling time $s_\ell$, the quantization center, $\hat{y}_{s_\ell - 1, \eta}$ is zero because of the deadbeat controller.
		This eliminates the need for running the controller at the encoder side and at the self-triggering module to compute the quantization center $\hat{y}_{s_\ell}$ for the standard observer-based controller, which is both computationally and implementation-wise more appealing.  
	\end{itemize}
Certainly, these two merits come at the price of a more complicated communication protocol between the S-C and C-A channels, increased communication overhead in the C-A channel as well as requiring a deadbeat controller gain matrix $K$.
The deadbeat condition \eqref{eq:dbdb} is more conservative compared with the Schur stable one, since only a few matrices $K$ obey $(\tilde{A} + \tilde{B} K)^\eta = 0$.
To increase the number of candidate matrices $K$, this condition can be relaxed by choosing $K$ such that $(\tilde{A} + \tilde{B} K)^i = 0$ with $i \in [\eta, n_x]$, and setting $\delta = \Delta/i$.
Under this condition, it has been shown in \cite{FahmyDead} that a large number of matrices $K$ can be constructed.  
This decreases the conservativeness of using the deadbeat controller.   
\end{remark}

\section{Self-triggered Control: DoS Case}\label{sec:predos}
Besides the communication cost, it has been reported that cyber-physical networked systems are often vulnerable to cyber-attacks, consisting of e.g., false-data injection attacks, switching attacks, replay attacks, and denial-of-service (DoS) attacks \cite{2013Attack,wu2020optimal,guo2023residual,Liu2022data}. 
For this reason, research on designing resilient control strategies against cyber-attacks has attracted lots of attention recently; see e.g., \cite{yuan2020resilient,chen2020Active,franze2021resilient}. 
In this section, we consider that the S-C channel is exposed to DoS attacks, and develop a DoS-resilient self-triggering mechanism and associated encoding scheme as well as stability analysis. 

\subsection{Denial-of-Service Attack}\label{sec:dos}
A brief introduction on the DoS attack is outlined.
Launched by adversaries, DoS attacks are intended to affect the timeliness of the information, and result in packet losses.
To characterize DoS attacks, we here employ the `duration-frequency' model initially studied in \cite{PersisInput} and later in
\cite{Persis2016Networked,8880482,FengResilient,Liu2021resilient}.

Before proceeding, let us define
at each time $s \in \mathbb{N}_{0}$ an attack-indicator function
\begin{equation}\label{eq:dosindicator}
h(s) := 
\begin{cases}
0,& {\text{no DoS attack at}}~s,\\
1,&{\text{DoS attack at}}~s.
\end{cases}
\end{equation}
For an arbitrary constant $\tau^a \in \mathbb{N}$, the DoS duration within the interval $[s, s + \tau^a)$ is defined as $\Phi_d(s, \tau^a) = \sum_{i = s}^{s + \tau^a - 1} h(i)$. 
Here, we only place an assumption on the DoS duration, since the result in this section demonstrates that the DoS frequency does not explicitly affect system stability in the considered setup.
\begin{assumption}[\emph{DoS duration}]\label{as:dosdur}
	There exist constants $\kappa_d \in \mathbb{R}_{\ge 0}$ and $\nu_d \in
	\mathbb{R}_{\ge 1}$, also known as chatter bound and average duration
	ratio, respectively, such that the DoS duration obeys
	\begin{equation}\label{eq:dosdur}
	\Phi_d(0,s) \le \kappa_d + {s}/{\nu_d}
	\end{equation}
	over the interval $[0,s)$ for all $s \in \mathbb{N}_{0}$.
\end{assumption}

\begin{remark}[\emph{Implication of DoS parameters}]
	Assumption \ref{as:dosdur} indicates that, the average duration of DoS attacks does not exceed a proportion $1/\nu_d$ of the entire interval.
	This assumption is general enough, since DoS attacks with higher strength can be modeled with a smaller $\nu_d$.
	In addition, as the  parameter $\nu_d$ approaches zero, this model characterizes a type of attacks preventing all packets from transmission.
	However, if no packet can be received successfully, then no controller can be constructed to stabilize open-loop unstable plants.
	To prevent such a situation, the condition $\nu_d \ge 1$ is placed. 
\end{remark}
\subsection{Resilient Control}
According to the previous subsection, no packet can be received at the decoder side if the current sampling time belongs to a DoS interval.
Therefore, our proposed self-triggering mechanism and encoding scheme for the DoS-free case cannot be directly employed and do not necessarily ensure system stability. 
In the following, we focus on the standard observer-based state feedback controller in Sec. \ref{sec:lq}.
The results can be generalized to the deadbeat controller \eqref{eq:dbctrl}.

Some assumptions are placed to show how the system reacts to the DoS attacks.
\begin{assumption}\label{as:zero}
	If there is a DoS attack at the transmission time $s_\ell$, i.e., $h(s_\ell) = 1$, then not quantized output $\mathcal{Q}(y_{s_\ell})$ is received, and a default zero will be used.
\end{assumption}
\begin{assumption}\label{as:ack}
	An ACK-based protocol is implemented by the S-C channel. 
	That is, the decoder sends an ACK to the encoder without delay when it receives a quantized output. 
\end{assumption}

 Assumption \ref{as:ack} enables the encoder to infer a DoS attack if no ACK is received. This assumption however can be removed if the deadbeat controller \eqref{eq:dbctrl} is adopted, which resembles that in \cite[Sec. IV]{Liu2021resilient}.
%
Under Assumption \ref{as:zero}, the observer-based state feedback controller  \eqref{eq:lqctrl} is replaced by
\begin{subequations}\label{eq:lqdosctrl}
	\begin{align}
\hat{x}_{s + 1} 	&=  A\hat{x}_{s}\! +\! \tilde{B} u_{s} \!+\!L[\mathcal{Q}(y_{s_\ell}) \!-\! \hat{y}_{s}],~ h(s_\ell)  = 0 \label{eq:lqdosctrl1}\\
	\hat{x}_{s + 1} 	&=  A\hat{x}_{s} + \tilde{B} u_{s}, ~~h(s_\ell)  = 1\label{eq:lqdosctrl2}\\
	\hat{y}_{s} 	&= C \hat{x}_{s} \label{eq:lqdosctrl3}\\
	u_{s} 	&= K \hat{x}_{s} \label{eq:lqdosctrl4}
	\end{align}
\end{subequations}
with the initial condition $\hat{x}_{0} = 0$.
Moreover, if a DoS attack occurs at a triggering time, that is, $h(s_\ell) = 1$, then the inequality \eqref{eq:lqtriinequ} is no longer valid.
To enhance system resilience against DoS attacks, we consider a mixed sampling scheme.
To be specific, i) during DoS attacks, the quantized output is communicated periodically with interval $\Delta$; and, ii) when there is no DoS attack, the next sampling time is calculated by \eqref{eq:lqselftri}.
In this manner, a successful transmission occurs as soon as an attack stops.
\begin{equation}\label{eq:lqdosselftri}
s_{\ell + 1} = 
\begin{cases}
\eqref{eq:lqselftri},~~& h(s_\ell) = 0\\
s_\ell + 1,~~ &h(s_\ell) = 1
\end{cases}.
\end{equation}
The update rule of $\{E_{s_\ell}\}_{\ell \in \mathbb{N}_{0}}$ in \eqref{eq:lqE} is modified as follows
\begin{equation}\label{eq:lqdosE}
E_{s_{\ell + 1}} :=
\begin{cases}
\eqref{eq:lqE}, & h(s_\ell) = 0\\
\omega_aE_{s_\ell}, &h(s_\ell) = 1 
\end{cases}
\end{equation}
where $\omega_a > 1$ is chosen such that $\Vert A^s\Vert_{\infty} \le (1/\Gamma) \omega_a^s$ holds for all $s \in \mathbb{N}_{0}$ with $\Gamma$ defined in \eqref{eq:gammaGamma}.
Based on \eqref{eq:lqdosctrl}--\eqref{eq:lqdosE}, our stability result comes as follows.
\begin{theorem}\label{thm:lqdos}
	Consider system \eqref{eq:dissys} with the controller \eqref{eq:lqdosctrl}, where matrices $L$ and $K$ are chosen such that $A - LC$ and $A + B K$ are Schur stable. 
	Under Assumptions \ref{as:abca}---\ref{as:initialx} and \ref{as:dosdur}---\ref{as:ack}, if  the sampling times $\{s_\ell\}_{\ell \in \mathbb{N}_{0}}$ are generated by \eqref{eq:lqdosselftri}, with i) the bound sequence $\{E_{s_{\ell}} \}_{\ell \in \mathbb{N}_{0}}$ in \eqref{eq:lqdosE}, ii) the threshold parameter ${\sigma} > 0$ obeying \eqref{eq:sigmacondition}, and iii) DoS attacks satisfying
	\begin{equation}\label{eq:doscondition}
	{1}/{\nu_d} < \big(\log(1/\omega_1)\big)/\big(\log(\omega_a/\omega_1)\big)
	\end{equation}
with $\omega_1$ defined in \eqref{eq:omega1}, 
	then the system \eqref{eq:dissys} is exponentially stable.
	In particular, the decay parameter $\omega  = \min\{\hat{\omega}_1, 1 \}$ satisfies Def. \ref{def:converge} for some constant $\Omega > 0$ where $\hat{\omega}_1 > \max \{\hat{\omega}, \omega_1(\omega_a/\omega_1)^{1/\nu_d} \}$ and $\hat{\omega}$ is defined in Thm. \ref{thm:lqconverge}.
\end{theorem} 
\begin{pf}
	We begin by constructing the sequence $\{E_{s_\ell}\}_{\ell \in \mathbb{N}_{0}}$ presented in \eqref{eq:lqdosE}.
	Based on Assumption \ref{as:zero}, if a DoS attack occurs at a sampling time, denoted by $s^{a} \in \mathbb{N}_{0}$, and lasts for $\tau^a$, then the error between the estimated state and the actual state obeys
	\begin{align}
		x_{s^{a} + i} - \hat{x}_{s^{a} + i} =   A_{d}^i(x_{s^{a}} - \hat{x}_{s^{a}}), ~i \in [0, \tau^a).
	\end{align}
	Based on the definition of $\omega_a$, and assuming that $\Vert x_{s^a} - \hat{x}_{s^a}\Vert_{\infty} \le E_{s^a}^x$ holds true, one gets that
	\begin{subequations}\label{eq:omega_a}
		\begin{align}
			\left\|x_{s^{a}+i}-\hat{x}_{s^{a}+i}\right\|_{A_{c l}} & \leq \Gamma\left\|x_{s^{a}+i}-\hat{x}_{s^{a}+i}\right\|_{\infty} \label{eq:omega_a1}\\
			& \leq \Gamma\Big(\frac{1}{\Gamma} \omega_{a}^{i}\left\|x_{s^{a}}-\hat{x}_{s^{a}}\right\|_{\infty}\Big) \label{eq:omega_a2}\\
			& \leq \omega_{a}^{i}\left\|x_{s^{a}}-\hat{x}_{s^{a}}\right\|_{A_{c l}} \label{eq:omega_a3}\\
			& \leq \omega_{a}^{i} E_{s^{a}}^{x}\label{eq:omega_a4}
		\end{align}
	\end{subequations}
	where equations \eqref{eq:omega_a1} and \eqref{eq:omega_a3} are derived following the definition of the norm $\Vert \cdot\Vert_{A_{cl}}$ in Lem. \ref{lem:norm}.
	Inequality \eqref{eq:omega_a2} follows from the definition of $\omega_a$ in \eqref{eq:lqdosE}.
	Let $E^{x}_{s^{a} + 1} := \omega_a E^{x}_{s^{a}}$, and hence $E_{s^{a} + 1} = \Vert C\Vert_{\infty} E^{x}_{s^{a} + 1} = \Vert C\Vert_{\infty}\omega_a E^{x}_{s^{a}}$.
	Therefore, sequence $\{E_{s_\ell}\}_{\ell \in \mathbb{N}_{0}}$ adheres to \eqref{eq:lqyE} with
	\begin{align*}
		\Vert y_{s^{a} + i} - \hat{y}_{s^{a} + i}\Vert_{\infty} &= \Vert C( x_{s^{a} + i} - \hat{x}_{s^{a} + i})\Vert_{\infty} \\
		&\le \Vert C\Vert_{\infty} E^{x}_{s^{a} + i} = E_{s^{a} + i}.
	\end{align*}	
	Capitalizing on the self-triggering mechanism in \eqref{eq:lqdosselftri} and the encoding scheme in \eqref{eq:lqdosE}, additional conditions on DoS attacks such that the system can achieve stability are derived in the following.
	
	
	For interval $[0, s_e)$ with arbitrary $s_e \in \mathbb{N}_{0}$, assume that there are $p \in \mathbb{N}_{0}$ DoS attacks launched within this interval; that is, there are $p$ DoS \emph{off/on} switches within this interval.
	To be specific, let $s^{a}_{m}$ and $\tau^{a}_{m}$ denote the beginning time and the duration time of the $m$-th DoS attack, respectively, where $m = 1, \cdots\!, p$.
	For simplicity, let $s^{a}_{1} > 0$ and $s^{a}_{p} + \tau^{a}_{p} < s_{e}$.
	Hence, it follows from the definition of the DoS duration that $\Phi_d(s_e) = \sum_{m = 1}^{p}\tau^{a}_{m}$.
	Consider that the adversary is aware of the self-triggering mechanism, and thus launches attacks at only the self-triggering times, namely $\left[s_{m}^{a}, s_{m}^{a}+\tau_{m}^{a}\right) \subset\left\{s_{\ell}\right\}_{\ell \in \mathbb{N}_{0}}$ for all $m =1, \cdots\!, p$.
	This can be the worst case, since each attack causes a packet loss and is effective.
	
	Since there is no attacks before $s^{a}_1$, one gets from \eqref{eq:lqdosE} that
	\begin{equation}
		E_{s^{a}_{1}}^{x} = \Gamma E_{in} \omega_1^{s^{a}_{1}}. 
	\end{equation}
	In addition, since no attacks happen within $[s^{a}_{m} + \tau^{a}_{m}, s^{a}_{m + 1}) \cup (s_p^a + \tau_p^a, s_e]$ for $m = 1, \cdots\!, p - 1$, it follows from \eqref{eq:lqdosE} that
	\begin{equation*}
		E_{s_{m+1}^{a}}^{x} \!\le \omega_{1}^{s_{m+1}^{a}-s_{m}^{a}-\tau_{m}^{a}} E_{s_{m}^{a}+\tau_{m}^{a}}^{x}\!=\! \omega_{1}^{s_{m+1}^{a}-s_{m}^{a}-\tau_{m}^{a}} \omega_{a}^{\tau_{m}^{a}} E_{s_{m}^{a}}^{x}
	\end{equation*}
	where $\omega_1 < 1$ is defined in \eqref{eq:omega1}.
	Combining the estimates \eqref{eq:eacl} and \eqref{eq:omega_a},
	we have from \eqref{eq:lqdosE} again that 
	\[
	\| x_{s_p^a} - \hat{x}_{s_p^a} \|_{A_{cl}} \leq E^x_{s_p^a}. 
	\]
		This implies that
		\begin{equation*}\label{eq:Espa}
			\Vert x_{s_p^a + \tau_p^a + q} - \hat{x}_{s_p^a + \tau_p^a + q} \Vert_{A_{cl}} \le \omega_1^{q}\omega_{a}^{\tau_{p}^{a}}E_{s_p^a}
		\end{equation*}
		for $q \in (0, s_e - s_p^a - \tau_p^a]$.
		Recursively,
		\begin{align*}
			\Vert e_{s_e}^i\Vert_{\infty} & = \Vert x_{s_e} - \hat{x}_{s_e}\Vert_{\infty}\\
			&\le \Gamma \omega_1^{s_e - \Phi_d(0,s_e)} \omega_a^{\Phi_d(0,s_e)} E_{in}\\
			&\le \Gamma\Big(\frac{\omega_a}{\omega_1}\Big)^{\kappa_d} \Big(\omega_1 \Big(\frac{\omega_a}{\omega_1}\Big)^{1/\nu_d}\Big)^{s_e} E_{in}\\
			&\le  \Gamma_2 \omega_3^{s_e} E_{in}
		\end{align*}
		where $\Gamma_2 := \Gamma(\omega_a/\omega_1)^{\kappa_d}$, and $\omega_3 := \omega_1 (\omega_a/\omega_1)^{1/\nu_d}$.
		Moreover, it follows from \eqref{eq:doscondition} that $\omega_3 <1$. 
		Therefore, the error between the actual state $x_{s_e}$ and the estimated output $\hat{x}_{s_e}$, i.e., $\Vert e_{s_e}^x \Vert = \Vert x_{s_e} - \hat{x}_{s_e}\Vert_{\infty}$, converges to the origin.
		
		Now, we are ready to derive the convergence bound for the state.
		Noticing from \eqref{eq:dissys} that
		\begin{subequations}
			\begin{align*}
				\Vert x_{s_e}\Vert_{\infty} &= \Big\Vert (A \!+\! B K)^{s_e}x_{0}\\
				& ~~~ -\sum_{i = 0}^{s_e - 1}(A \!+\! B K)^{s_e - i - 1}B K e^{x}_{i}\Big\Vert_{\infty}\nonumber\\
				& \le \Vert (A \!+\! B K)^{s_e}\Vert_{\infty}\Vert x_{0}\Vert_{\infty} \\
				&~~~ +  \sum_{i = 0}^{s_e - 1}\Vert(A \!+\! B K)^{s_e - i - 1}B K\Vert_{\infty} \Vert e^{x}_{i}\Vert_{\infty}\nonumber.
			\end{align*}	
		\end{subequations}	
		Combining with \eqref{eq:omega12}, similar from \eqref{eq:lqxconverge} in the proof of Thm. \ref{thm:lqconverge}, we have that 
		\begin{subequations}
			\begin{align}
				\Vert x_{s_e}\Vert_{\infty} &\le \Omega_K E_{in} \omega^{s_e} \nonumber\\
				&~~~+ \underbrace{(\Vert B K\Vert_{\infty} \Omega_K \Gamma_2)}_{\Omega_4} E_{in}\sum_{i = 0}^{s_e - 1}\omega^{s_e - i - 1}\omega_3^{i}\nonumber\\
				& \le \Omega_K E_{in} \omega^{s_e} + \Omega_4 E_{in} \sum_{i = 0}^{s_e - 1} \omega_4^{s_e - 1}\\
				&\le \Omega_K E_{in} \omega^{s_e} + \Omega_4 E_{in} (s_e - 1)\omega_4^{s_e}/\omega_4 \\
				&\le  \Omega_5 E_{in}\hat{\omega}_4^{s_e}
			\end{align}
		\end{subequations} 
		where $\omega_4 := \max\{\omega, \omega_3\}$, $\hat{\omega}_4 > \max\{\omega, \omega_4\}$ and $\Omega_5 = \Omega_K + \Omega_4 c_6/\hat{\omega}_4$ with $(s_e - 1)\omega_4^{s_e - 1}\le  c_6 \hat{\omega}_4^{s_e - 1}$.
		This completes the proof according to Def. \ref{def:converge}.
	\end{pf}	
\begin{remark}[\emph{Conservativeness of DoS bound}]\label{rmk:dosconservative}
	Condition \eqref{eq:doscondition} is derived by only taking into account the DoS attacks at self-triggering times.
	Since attacks that occur between two consecutive self-triggering times $
	(s_\ell, s_{\ell + 1})$ are ignored, this bound could be the worst, where the adversary knows the self-triggering mechanism.
	Therefore, to save energy during inter-sampling times, attacker launches attacks only at the sampling times.
	Hence, if more information on DoS attacks is provided, less conservative bounds can be obtained.
\end{remark}
\begin{remark}
	{\bf {\rm \bf(}\emph{Connection with the work \cite{8880482}}{\rm \bf)}}\!
	Condition \eqref{eq:doscondition} indicates that under the proposed encoding schemes and self-triggering mechanism, system achieves exponential stability without any DoS frequency assumption.
	In addition, instead of \eqref{eq:lqdosE}, if we let i) $E_{s + 1} = \omega_1 E_{s}$ for $h(s) = 0$, and ii) $E_{s + 1} = \omega_a E_{s}$ for $h(s) = 1$, then the bound sequence $\{E_{s_\ell}\}_{\ell \in \mathbb{N}_{0}}$ is updated periodically.
	In this setting, the condition \eqref{eq:doscondition} coincides with the one derived in \cite{8880482}.
	Moreover, to enhance the system resilience against DoS attacks, one can choose smaller $\sigma$, which results in smaller inter-sampling times and more samplings.
\end{remark}

%

	\section{Numerical Example}
	\label{sec:simulation}
	To numerically test the performance of the proposed self-triggered quantized controllers, we call for the unstable batch reactor in \cite{Walsh2002Scheduling}, for which a linearized model has been studied in e.g., \cite{8880482,Liu2021resilient}.
	The system matrices are given by
	\begin{align*}
	A :=
	\left[
	\begin{matrix}
	1.38 & -0.2077 & 6.715 & -5.676\\
	-0.5814 & -4.29 & 0 & 0.675\\
	1.067 & 4.273 & -6.654 & 5.893\\
	0.048 & 4.273 & -1.343 & -2.104
	\end{matrix}
	\right]\\
	B := \left[
	\begin{matrix}
	0 & 0\\
	5.679 & 0\\
	1.136 & -3.146\\
	1.136 & 0
	\end{matrix}
	\right],
	C :=
	\left[
	\begin{matrix}
	1 & 0 & 1 & -1\\
	0 & 1 & 0 & 0
	\end{matrix}
	\right].
	\end{align*}
	The system $(A, B, C)$ is observable and controllable with $\eta =  2$.
	For the feedback controller \eqref{eq:lqctrl} in Sec. \ref{sec:lq}, we discretize the system with a sampling period $\Delta = 0.005$; and for the observer-based deadbeat controller \eqref{eq:dbctrl} in Sec. \ref{sec:db}, we have $\delta = \Delta/\eta = 0.005$.
	In this setting, the C-A channel communication costs incurred by the controllers \eqref{eq:lqctrl} and \eqref{eq:dbctrl} are the same.
	Choose the gain matrices $K_1$ and $K_2$, such that $A + BK_1$ is Schur Stable and that $(A + BK_2)^\eta = 0$ holds, yielding
	\begin{align*}
	&K_1 := \left[
	\begin{matrix}
	 1.4110 &  -3.5708 &  -0.6385 &  -4.1134\\
	6.0726  & -0.0486  &  4.6801  & -2.5005
	\end{matrix}
	\right]\\
	& K_2 :=
	\left[
	\begin{matrix}
	288.3  &  233.4 &   1000.0 &  -1429.4\\
	1945.2 &  -1.9  &  94.7 &  -84.7
	\end{matrix}
	\right].
	\end{align*}
	
	The gain observers $L$ and $M$ are chosen such that $A -LC$ and $\tilde{A}^\eta (I - MC)$ are both Schur stable.
		Based on \eqref{eq:sigmacondition} and \eqref{eq:dbsigma}, take the parameters of the self-triggering mechanism to be $\sigma = 0.0343$, and the maximum inter-sampling time to be $\tau_{\max} = 20$.
		\begin{figure}[htbp]
			\centering
			\includegraphics[width=8.5cm]{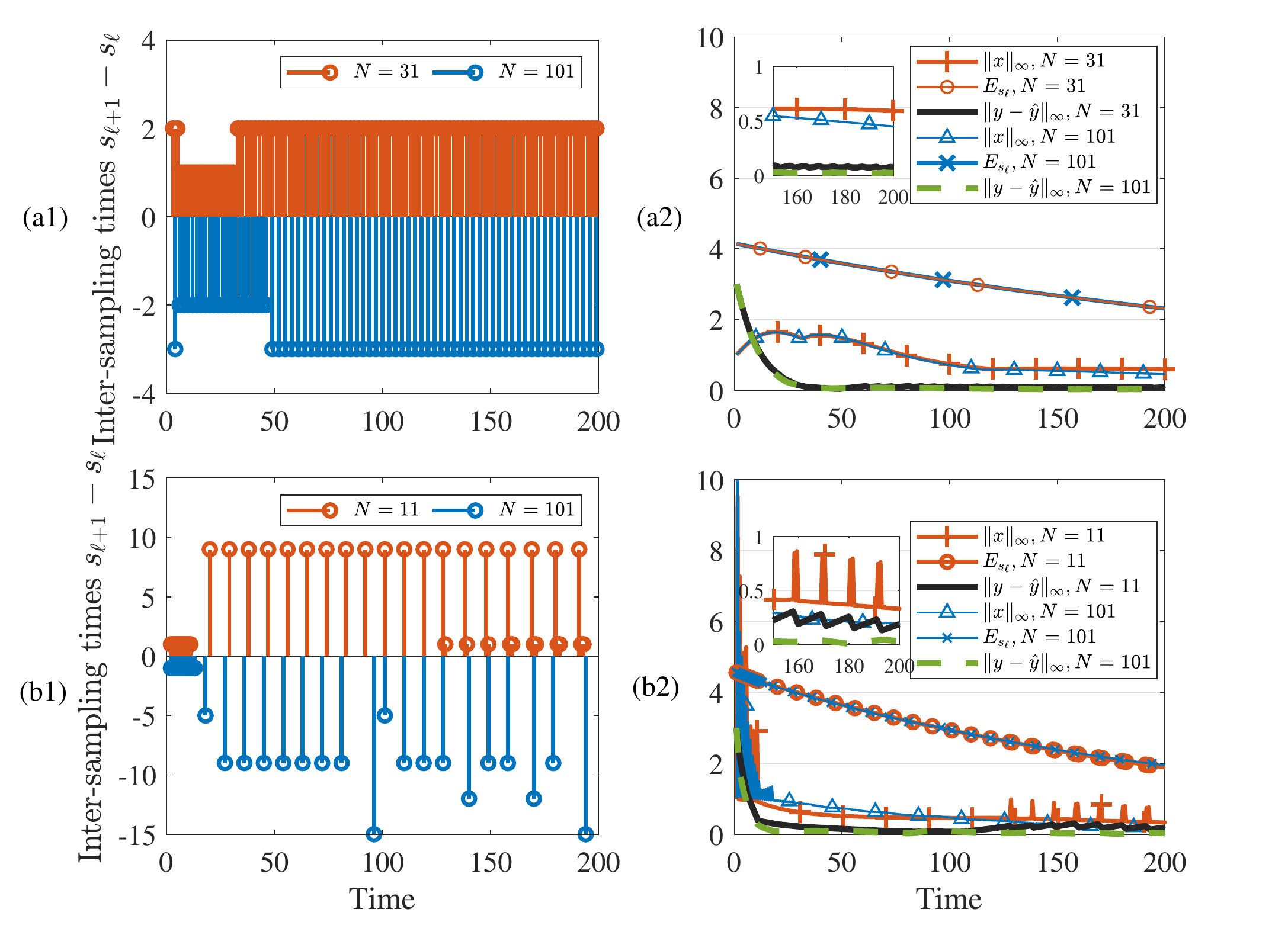}\\
			\caption{System performance with different controllers and different quantization levels. 
				Panels (a1) and (a2): controller \eqref{eq:lqctrl} under $N = 31$ and $N = 101$.
				Panels (b1) and (b2): controller \eqref{eq:dbctrl} under $N = 11$ and $N = 101$.
				 }\label{fig:lqdb_N}
			\centering
		\end{figure}
	\begin{figure}[htbp]
		\centering
		\includegraphics[width=8.5cm]{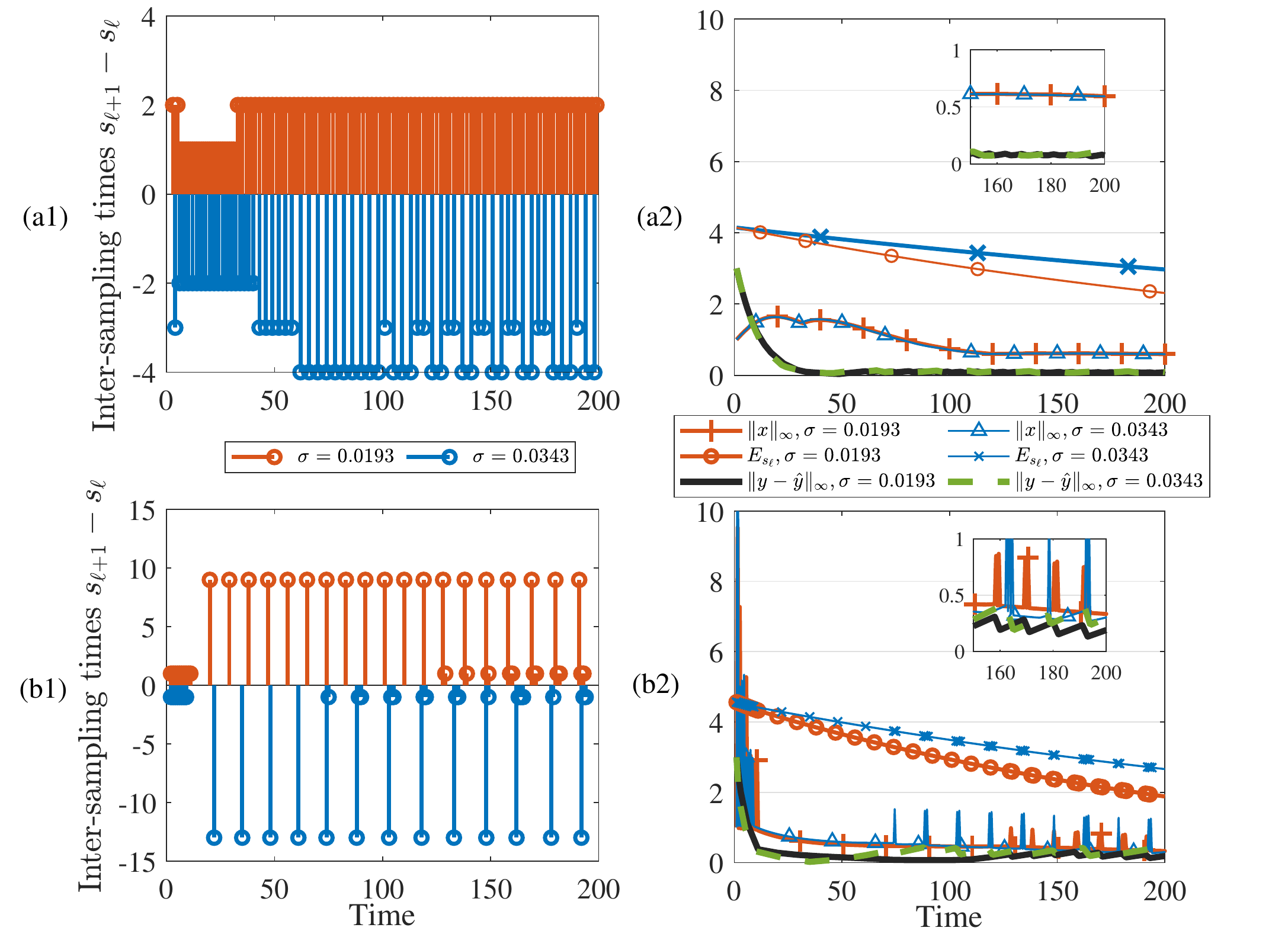}\\
		\caption{System performance under different $\sigma$.
			Panels (a1) and (a2): controller \eqref{eq:lqctrl} under $N = 31$ with $\sigma = 0.0193$ and $\sigma = 0.0343$.
			Panels (b1) and (b2): controller \eqref{eq:dbctrl} under $N = 11$ with $\sigma = 0.0193$ and $\sigma = 0.0343$.}
		\label{fig:lqdb_sigma}
		\centering
	\end{figure}

		First, system performance in the DoS-free case using different controllers under different levels of quantization  is reported in Fig. \ref{fig:lqdb_N}.
It is evident in \eqref{eq:lqg} that a larger quantization level results in a larger inter-sampling time and thus less samples.
However, in Fig. \ref{fig:lqdb_N}, change in the quantization level has only little effect on the convergence of state.
In addition, compared with the standard controller, system under the deadbeat controller not only incurs less samplings but also converges in a faster rate.

		For a given quantization level, Fig. \ref{fig:lqdb_sigma} compares the performance under different values of $\sigma$.
		It is clear that smaller $\sigma$ leads to faster convergence rate of the quantization range $E$, whereas yielding smaller inter-sampling interval $s_{\ell + 1} - s_\ell$.
		This can be observed from the definition of $\omega_1$ in \eqref{eq:omega1}.
\begin{figure}[htbp]
	\centering
	\includegraphics[width=8.5cm]{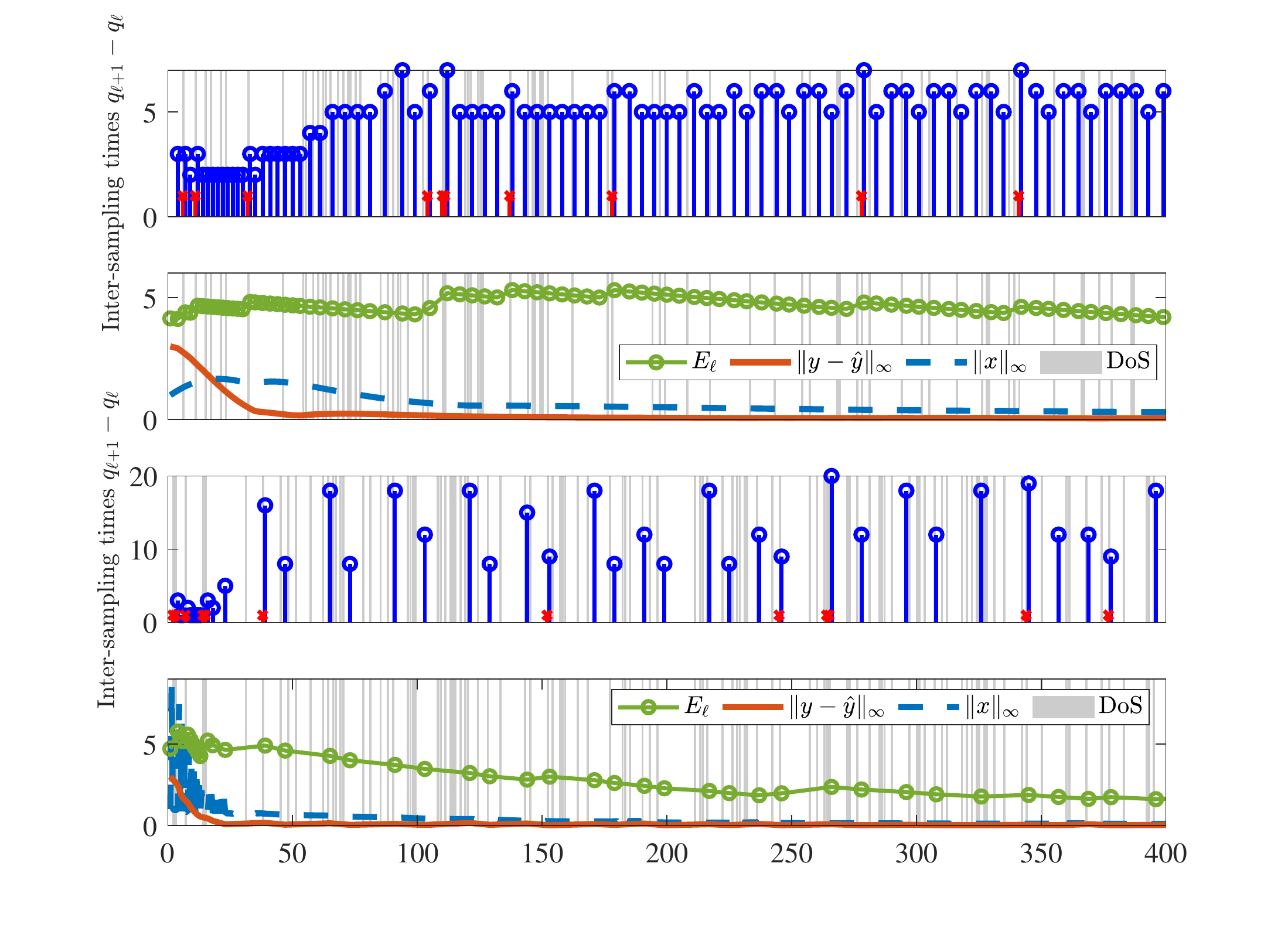}\\
	\caption{System performance with controller under DoS attacks with $N = 101$: controller \eqref{eq:lqctrl} (top two) vs. controller \eqref{eq:dbctrl} (bottom two).}\label{fig:lqN101dos}
	\centering
\end{figure}

		Moreover, consider DoS attacks at the S-C channel, which are generated randomly and represented by gray shades.
		The red cross marked lines represent the sampling times that fail due to DoS attacks.	
		Set the quantization level $N = 101$.
		Fig. \ref{fig:lqN101dos} depicts the system performance under DoS attacks with controllers in Secs. \ref{sec:lq} (top two) and \ref{sec:db} (bottom two).
		Collecting only the effective DoS attacks, denoted by $\Phi_d^e(0,s)$ (i.e., attacks that happen at sampling times, represented by red cross marked line) in the top panel of Fig. \ref{fig:lqN101dos}, we have that $\Phi_d^e(0,400) = 10$.
		Let $\kappa_d = 1$ and $\Phi_d^e(0,400) = 10$ obey Assumption \ref{as:dosdur} by setting $\nu_d = 44$.
		Similarly, for the bottom panel in Fig. \ref{fig:lqN101dos}, the number of effective DoS attacks is $\Phi_d^e(0,400) = 12$, and Assumption \ref{as:dosdur} is met with $\kappa_d = 1$ and $\nu_d = 36$.
		In addition, according to Thm. \ref{thm:lqdos}, system with controller \eqref{eq:lqctrl} achieves stability if DoS attacks obey $1/\nu_d \le 0.0042$ ($\nu_d \ge 238$) and system adopts the deadbeat controller in \eqref{eq:dbctrl} achieves stability if DoS attacks obey $1/\nu_d \le 0.0056$ ($\nu_d \ge 179$).
		This reveals the conservativeness of the result \eqref{eq:doscondition}.
\begin{figure}[htbp]
	\centering
	\includegraphics[width=8.5cm]{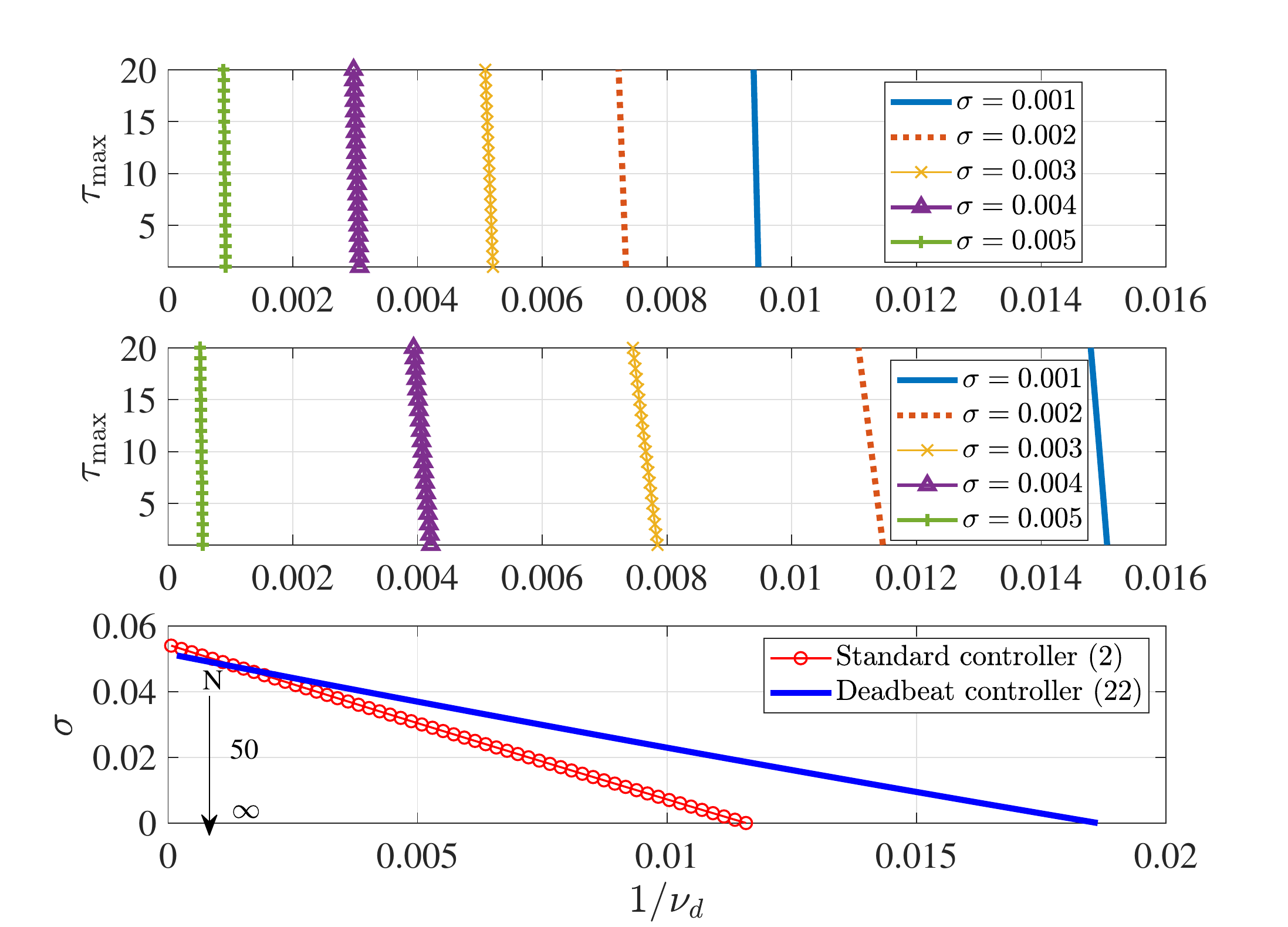}\\
	\caption{Relationship between the self-triggering parameters and DoS duration. The maximum inter-sampling time $\tau_{\max}$: controller \eqref{eq:lqctrl} (top) vs. controller \eqref{eq:dbctrl} (middle).
	}\label{fig:sigma_dos}
	\centering
\end{figure}	
Finally, the relationship between the self-triggering parameters $\sigma$, $\tau_{\max}$, and the upper bound on DoS duration $1/\nu_d$ under which the state converges, is depicted in Fig. \ref{fig:sigma_dos}. 
	It can be seen that to achieve better resilience against DoS attacks, one can choose smaller $\sigma$ (see the bottom panel in Fig. \ref{fig:sigma_dos}), and the choice of maximum inter-sampling time has minimal influence on the system resilience (see the top two panels in Fig. \ref{fig:sigma_dos}). 
	In addition, the bottom panel in Fig. \ref{fig:sigma_dos} illustrates that under the same $\sigma$, system with controller \eqref{eq:dbctrl} achieves better resilience against DoS attacks while requiring less samplings.
	Furthermore, \eqref{eq:sigmacondition} and \eqref{eq:dbsigma} indicate that the lower bound of $\sigma$ depends on $N$.
	This is depicted in Fig. \ref{fig:sigma_dos} by the axis $N$.
	As $N\rightarrow \infty$, the lower bound of $\sigma$ approaches zero, and one can choose smaller $\sigma$ to enhance system resilience against DoS attacks.
	\section{Conclusions}\label{conclusion}
This paper investigated the self-triggered stabilization problem of linear control systems with quantized output. 
Employing a standard observer-based state feedback control law, 
a self-triggering mechanism and an output encoding scheme were designed such that exponential system stability is achieved. When faster communication rates can be afforded at the C-A channel, an observer-based deadbeat controller was developed which can considerably simplify the self-triggering mechanism as well as save the computational and communication resources.
	Indeed, the resulting self-triggering mechanism for quantized output control is as simple as that for quantized state feedback control.
	To further endow the system with resilience against DoS attacks, some modifications were made to the proposed self-triggering mechanism and to the output encoding scheme.
	It was shown that as long as the DoS duration time does not exceed a certain bound, the closed-loop system is guaranteed to be exponentially stable.
	Moreover, a trade-off was revealed between the maximum inter-sampling time and the resilience.
Finally, numerical tests were provided to showcase the effectiveness of the proposed self-triggered quantized controllers.

This article also opens up several avenues for future research.
To name a few, since the self-triggering parameters are related with the controller gain $K$, the results in Sec. \ref{sec:lq} can be extended by using a model predictive controller.
Moreover, for the deadbeat controller in Sec. \ref{sec:db}, designing an optimal observer gain $L$ to further reduce the required samplings is another interesting direction.

	\bibliographystyle{plainnat}
	\bibliography{bible3}

\end{document}